\documentclass[%
 reprint,
superscriptaddress,
preprintnumbers,
 amsmath,amssymb,
 aps,
 twocolumn,
]{revtex4}
\pdfoutput=1

\usepackage{amsmath}
\usepackage{graphicx}
\usepackage{bbold}
\usepackage{multirow}
\usepackage{rotating}

\usepackage{color}
\newcommand{\rnu}{\langle r_\nu^2 \rangle}

\usepackage[svgnames]{xcolor}
\usepackage{hyperref}% add hypertext capabilities
\hypersetup{colorlinks=True,
					breaklinks=True,
             		urlcolor=RoyalBlue,
             		citecolor=RoyalBlue,
             		linkcolor=RoyalBlue}

\begin{document}

\preprint{IFIC/19-45,YITP-SB-19-37}

\title{Coherent Elastic Neutrino-Nucleus Scattering at the European Spallation Source}

\author{D.~Baxter}
  \affiliation{Enrico Fermi Institute, Kavli Institute
    for Cosmological Physics, and Department of Physics
University of Chicago, Chicago, Illinois 60637, USA} 
\author{J.I.~Collar}
\email{collar@uchicago.edu}
  \affiliation{Enrico Fermi Institute, Kavli Institute
    for Cosmological Physics, and Department of Physics
University of Chicago, Chicago, Illinois 60637, USA} 
\author{P.~Coloma}
\email{pcoloma@ific.uv.es}
\affiliation{
 Instituto de F\'{\i}sica Corpuscular, Universitat de Val\'encia and CSIC,
 Edificio Institutos Investigaci\'on, Catedr\'atico Jos\'e Beltr\'an 2,
 46980 Valencia, Spain}  
\author{C.E.~Dahl}
\affiliation{Department of Physics and Astronomy, Northwestern University, Evanston, Illinois 60208, USA}
\affiliation{Fermi National Accelerator Laboratory, Batavia, Illinois60510, USA}
\author{I.~Esteban}
\email{ivan.esteban@fqa.ub.edu}
\affiliation{Departament  de  Fisica  Quantica  i  Astrofisica
 and  Institut  de  Ciencies  del  Cosmos,  Universitat
 de Barcelona, Diagonal 647, E-08028 Barcelona, Spain}
\author{P.~Ferrario}
\email{paola.ferrario@dipc.org}
\affiliation{Donostia International Physics Center (DIPC),
  Paseo Manuel Lardizabal, 4, Donostia-San Sebasti\'an,
E-20018, Spain}
\affiliation{Ikerbasque, Basque Foundation for Science, Bilbao, E-48013, Spain}
\author{J.J.~Gomez-Cadenas}
\email{jjgomezcadenas@dipc.org}
\affiliation{Donostia International Physics Center (DIPC),
  Paseo Manuel Lardizabal, 4, Donostia-San Sebasti\'an,
E-20018, Spain}
\affiliation{Ikerbasque, Basque Foundation for Science, Bilbao, E-48013, Spain}
\author{M.~C.~Gonzalez--Garcia}
\email{maria.gonzalez-garcia@stonybrook.edu}
\affiliation{Departament  de  Fisica  Quantica  i  Astrofisica
 and  Institut  de  Ciencies  del  Cosmos,  Universitat
 de Barcelona, Diagonal 647, E-08028 Barcelona, Spain}
\affiliation{Instituci\'o Catalana de Recerca i Estudis Avancats (ICREA)
Pg. Lluis  Companys  23,  08010 Barcelona, Spain.}
\affiliation{C.N. Yang Institute for Theoretical Physics, Stony Brook University, Stony Brook NY11794-3849,  USA}
\author{A.R.L.~Kavner}
  \affiliation{Enrico Fermi Institute, Kavli Institute
    for Cosmological Physics, and Department of Physics
University of Chicago, Chicago, Illinois 60637, USA} 
\author{C.M.~Lewis}
  \affiliation{Enrico Fermi Institute, Kavli Institute
    for Cosmological Physics, and Department of Physics
University of Chicago, Chicago, Illinois 60637, USA} 
\author{F.~Monrabal}
\email{francesc.monrabal@dipc.org}
\affiliation{Donostia International Physics Center (DIPC),
  Paseo Manuel Lardizabal, 4, Donostia-San Sebasti\'an,
E-20018, Spain}
\affiliation{Ikerbasque, Basque Foundation for Science, Bilbao, E-48013, Spain}
\author{J.~Mu\~noz Vidal}
\affiliation{Donostia International Physics Center (DIPC),
  Paseo Manuel Lardizabal, 4, Donostia-San Sebasti\'an,
E-20018, Spain}  
\author{P.~Privitera}
  \affiliation{Enrico Fermi Institute, Kavli Institute
    for Cosmological Physics, and Department of Physics
University of Chicago, Chicago, Illinois 60637, USA} 
\author{K.~Ramanathan}
  \affiliation{Enrico Fermi Institute, Kavli Institute
    for Cosmological Physics, and Department of Physics
University of Chicago, Chicago, Illinois 60637, USA} 
\author{J.~Renner}
\affiliation{Instituto Gallego de F\'isica de Altas Energ\'ias, Univ.\ de Santiago de Compostela, 
 Campus sur, R\'ua Xos\'e Mar\'ia Su\'arez N\'u\~nez, s/n, Santiago de Compostela, E-15782, Spain}

%\date{\today}

\begin{abstract}
The European Spallation Source (ESS), presently well on its way to
completion, will soon provide the most intense neutron beams for
multi-disciplinary science. Fortuitously, it will also generate the
largest pulsed neutrino flux suitable for the detection of Coherent
Elastic Neutrino-Nucleus Scattering (CE$\nu$NS), a process recently
measured for the first time at ORNL's Spallation Neutron Source. We
describe innovative detector technologies maximally able to profit
from the order-of-magnitude increase in neutrino flux provided by the
ESS, along with their sensitivity to a rich particle physics
phenomenology accessible through high-statistics, precision CE$\nu$NS
measurements. \end{abstract}

\maketitle

%\tableofcontents

\section{Introduction}
\label{sec:intro}
Low-energy neutrinos can scatter off the atomic nucleus as a whole,
via the weak neutral current. During this process the initial and
final states of the nuclear target are indistinguishable, permitting a
coherent contribution from all nucleons. The net result is a drastic
enhancement to the cross-section for this type of neutrino
interaction, roughly proportional to the square of the number of
neutrons present in the target nucleus. The single observable from
this so-called Coherent Elastic Neutrino-Nucleus Scattering
(CE$\nu$NS) is a recoiling nucleus, which generates a signal in the
few keV to sub-keV energy range, difficult to reach with most
contemporary radiation detectors. An additional obstacle to CE$\nu$NS
detection is the limited number of suitable neutrino sources, both
sufficiently intense in yield, and low enough in neutrino energy so
the coherence condition can be satisfied (that is, $|Q| < 1/R$, where
$|Q|$ is the momentum transfer and $R$ is the radius of the nucleus).  

Perhaps the most convenient of the available possibilities are the
neutrinos produced following the decay at rest of positive pions at
spallation sources. The first advantage these provide is the
generation of nuclear recoils as energetic as is allowed by the coherence condition,
facilitating their detection \cite{leo}. Additionally, the pulsed beam
timing that is typically characteristic of this type of source reduces
the impact of steady-state backgrounds able to mask the signal, in
proportion to a small duty factor. CE$\nu$NS was experimentally
demonstrated by the COHERENT experiment \cite{science} forty-three
years following its theoretical description \cite{freedman}, using the
presently most intense Spallation Neutron Source (SNS), sited at the
Oak Ridge National Laboratory, USA. A low-background 14.6 kg CsI[Na]
scintillator with optimal characteristics for this goal
\cite{ournim,bjorn} was employed as the detecting medium.

Besides a miniaturization of neutrino detectors, and any future
technological applications that might bring, CE$\nu$NS has been
proposed as a new tool for the study of fundamental neutrino
properties. Phenomenological work following the first CE$\nu$NS
measurement has generated improved bounds on non-standard neutrino 
interactions (NSI)
~\cite{nsi2,nsi1,nsi3,nsi4,nsi5,nsi6,nsi7,nsi8,nsi9,Giunti:2019xpr,Denton:2018xmq},
as well as contributions to the study of nuclear
structure~\cite{nst1,nst2,nst3,nst4,Cadeddu:2019eta}. The possibility
of constraining our present knowledge of neutrino electromagnetic
properties \cite{em1,em2,em3,em4,Papoulias:2019txv} and of the weak
mixing angle \cite{wma1,wma2,wma3}, together with the potential to
search for sterile neutrinos \cite{ste1,carlos}, or for new types of
dark matter particles \cite{dm1,dm2,dm3}, has also been examined.

\begin{figure}[!htbp]
\centering
\includegraphics[width=1.04\columnwidth]{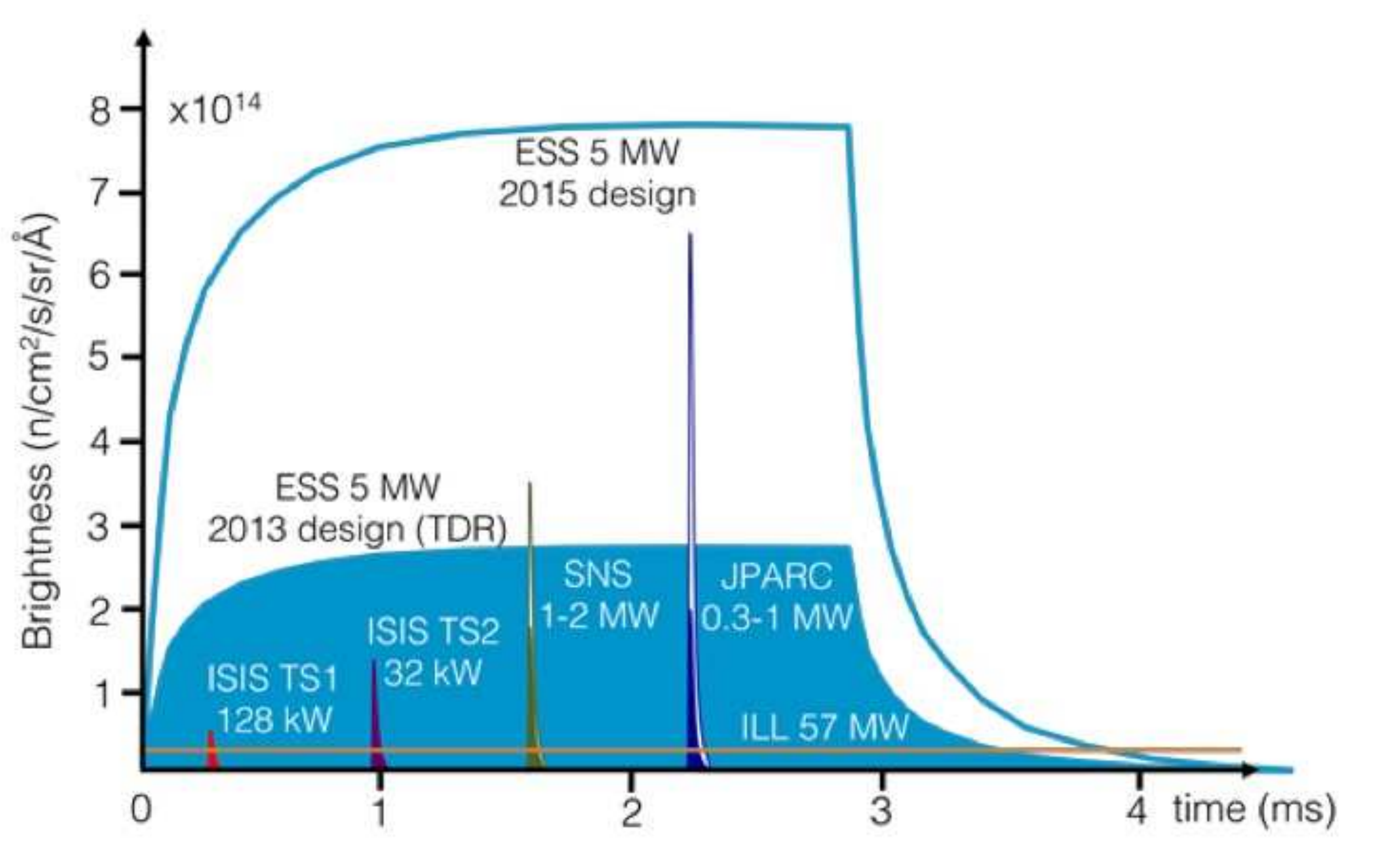} 
\caption{\label{fig:ess}(Source: ESS) Neutron production from existing
  and planned spallation sources. The nominal SNS power is 1 MW at
  proton energy 1 GeV, with a plan to reach 2 MW by 2026. The ESS
  power will be 5 MW at 2 GeV circa 2023, with the ability to further
  upgrade. Differences in the duration of the protons-on-target (POT)
  pulse are visible in the figure. The ESS will generate an increase
  in neutron brightness by a factor 30-100 with respect to previous
  spallation sources, and an order of magnitude larger neutrino yield
  than the SNS.}
\end{figure}

In this work we describe the opportunity to perform high-statistics
CE$\nu$NS measurements provided by the upcoming European Spallation
Source (ESS) sited in Lund, Sweden. The ESS will combine the world's
most powerful superconducting proton linac with an advanced hydrogen
moderator, generating the most intense neutron beams for
multi-disciplinary science (Fig.\ \ref{fig:ess}). It will also provide
an order of magnitude increase in neutrino flux with respect to the
SNS. This will facilitate CE$\nu$NS measurements not limited in their
sensitivity to new physics (NP) by poor signal statistics, while still
employing non-intrusive, compact (few kg) neutrino detectors, able to
operate without interference with ESS neutron activities.

%\begin{widetext}
%\end{widetext}

This manuscript is organized as follows. Section~\ref{sec:ESSvsSNS}
describes the
characteristics of the ESS as a neutrino source for CE$\nu$NS,
establishing a positive comparison with the SNS in all aspects
involved, while also delineating the ESS site characterization
activities that will be necessary to confirm this strong
potential. Section~\ref{sec:detectors}
briefly describes a number of state-of-the-art
nuclear recoil detector technologies maximally able to exploit the
opportunity that the ESS represents. Section~\ref{sec:pheno}
discusses the physics
reach provided by the combination of this source and these detectors,
on a number of phenomenological fronts probing for deviations from the
Standard Model {(SM)}. Our conclusions are  presented in
Section~\ref{sec:conclusions}. Chief among those is the unique opportunity provided by the ESS to
perform {\it precision} studies of CE$\nu$NS, for which the statistics
of the neutrino signal will contribute a subdominant uncertainty. 

\section{The ESS as a neutrino source: comparison to the SNS}
\label{sec:ESSvsSNS}

At spallation sources, both $\pi^+$ and $\pi^-$ are produced in
proton-nucleus collisions in the target. While the resulting $\pi^-$
are efficiently absorbed by nuclei before they can decay, the produced
$\pi^+$ lose energy as they propagate in the target and will
eventually decay at rest (DAR) into $\pi^+ \to \mu^+ \nu_\mu$,
followed in close spatial vicinity (within $\sim$0.2 g/cm$^{2}$) by
$\mu^+ \to e^+ \nu_e \bar\nu_\mu$. Three neutrino flavors with
essentially identical CE$\nu$NS cross section \cite{diff}, are
therefore engendered for each $\pi^+$ created. Being the result of a
two-body decay, the $\nu_\mu$ flux is monochromatic: $E_{\nu_\mu} =
(m_\pi^2 - m_\mu^2)/(2 m_\pi) \simeq 29.7$~MeV, where $m_\pi$ and
$m_\mu$ refer to the pion and muon masses, respectively. Conversely,
the $\nu_e$ and $\bar\nu_\mu$ fluxes follow a continuous distribution
at energies $E_{\nu_e, \bar\nu_\mu} < m_\mu/2\simeq
52.8$~MeV. Normalized to one, they read:
\begin{align}
f_{\bar\nu_\mu}(E_\nu) & = \frac{64}{m_\mu}\left[\left(\frac{E_\nu}{m_\mu}\right)^2\left(\frac34-\frac{E_\nu}{m_\mu}\right)\right]\,,\label{eq:COHflux} \\
f_{\nu_e}(E_\nu) & = \frac{192}{m_\mu}\left[\left(\frac{E_\nu}{m_\mu}\right)^2\left(\frac12-\frac{E_\nu}{m_\mu}\right)\right]\, . \nonumber 
\end{align}
Since the lifetime of the muon is much longer than that of the pion, the
monochromatic component is usually referred to as the prompt
contribution to the flux, as opposed to the delayed contributions from
$\mu^+$ decay. 
%%%%%%%%%%%%%%%%%%%%%%%%%%
\begin{figure}
    \centering
    \includegraphics[width=1.0\columnwidth]{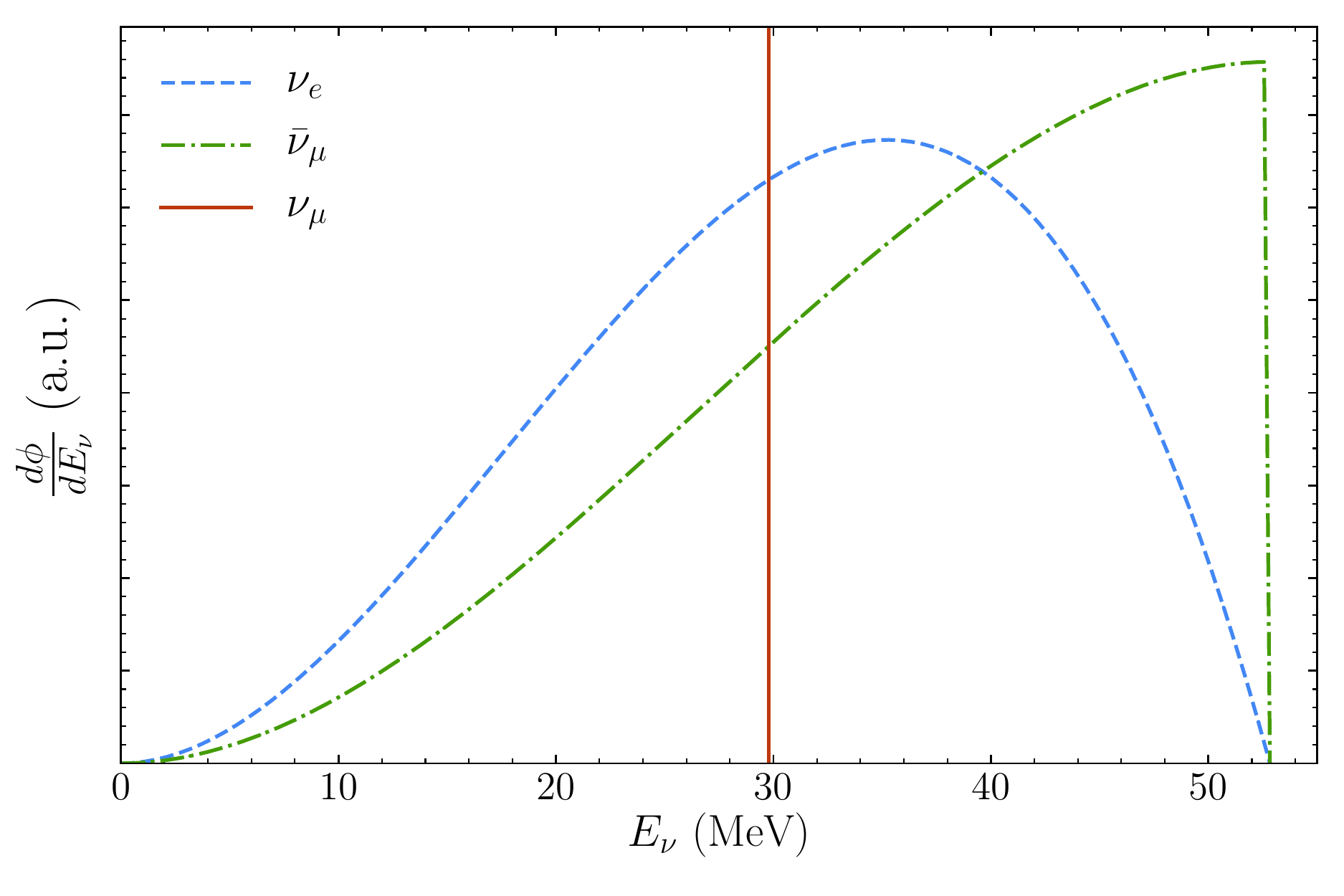}
    \caption{Neutrino flux spectra expected from pion DAR, in
      arbitrary units (a.u.), as a function of the neutrino energy in
      MeV. The three components of the flux are shown separately as
      indicated by the legend. The distributions have been normalized
      to one. }
    \label{fig:spectra}
\end{figure}
%%%%%%%%%%%%%%%%%%%%%%%%%%
For reference, the flux spectra is shown in Fig.~\ref{fig:spectra},
for the three components separately.

Besides the obvious gain in statistics with respect to other neutrino
sources, the use of neutrinos from pion decay at rest presents a clear
advantage: the energy dependence of the flux is well-known in this
case, and there is only room for systematic uncertainties affecting
its normalization. This contrasts with the large uncertainties
associated to conventional neutrino beams, where neutrinos are
produced in the decay of pions and kaons in flight, and the
determination of the neutrino spectral shape relies on Monte Carlo
simulations.

The described energy spectrum for neutrinos from spallation sources is
rather independent from the characteristics of the proton beam, and
therefore it is very similar for SNS and ESS.  There are however,
important differences in intensity and temporal structure that we
describe next.

The ESS is scheduled to reach its design power of 5 MW and goal proton
energy of 2 GeV by 2023, with a neutron user program commencing soon
after \cite{essdesign}. First protons-on-target (POT) are imminent,
expected for 2021 at a reduced 0.5 GeV. The power delivered by a
spallation source can be regarded as the product of proton current,
and energy per proton. From this perspective, the proton energy of
$\sim$1 GeV and nominal 1 MW design power of the SNS implies an
increase in average proton current at the ESS by a factor of 2.5, for
a total of 2.8$\times 10^{23}$ POT per calendar year. Both ESS and SNS
operate on a scheduled 5,000 hours of beam delivery per year, the
downtime being reserved for facility and accelerator maintenance.

The decay-at-rest (DAR) neutrino yield is expected to increase rapidly
with proton energy at spallation sources. A dedicated calculation of
the ESS neutrino yield \cite{burman,report} modified the LAHET
(Bertini model) Monte Carlo code \cite{lahet} to include experimental
data on pion production in the 0.5-2.5 GeV proton energy range. This
reference work { found} an increase by a factor 4.4 in neutrino yield for
a tungsten target, in going from the nominal 0.94 GeV proton energy of
the SNS \cite{science} to the 2 GeV of a completed ESS. In particular,
the probability of inducing a second pion-generating nuclear
interaction per initial proton is predicted to climb rapidly with
proton energy, as does double-pion production per interaction, and the
ratio of pion decays to captures \cite{burman}. A comparison of
theoretical predictions with experimental data for a number of
neutrino cross section measurements assigns a modest uncertainty to
these DAR neutrino production calculations, validating them
\cite{louis}.

\begin{figure}[!htbp]
  \centering
\includegraphics[width=1.0\columnwidth]{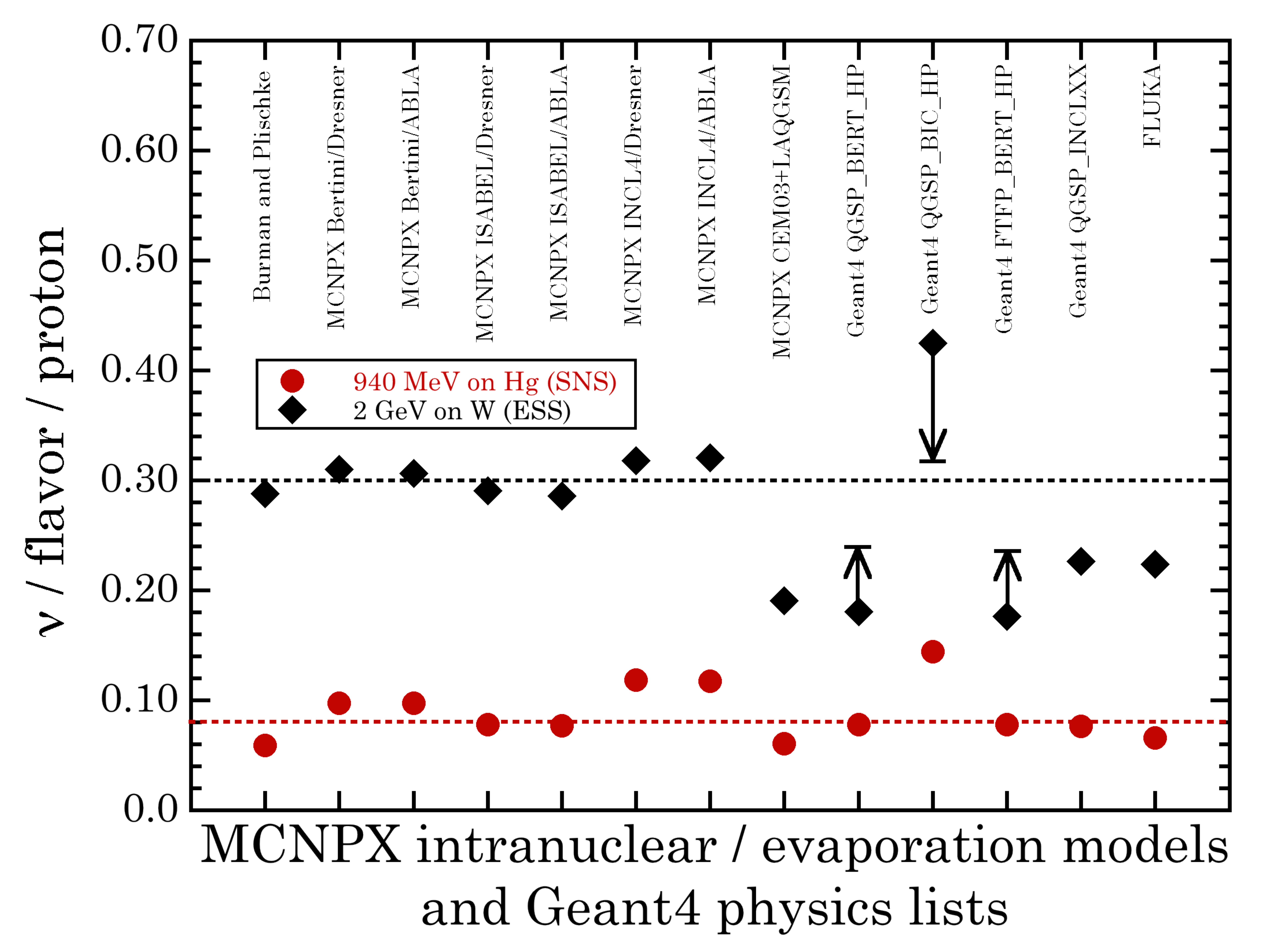}   
\caption{\label{fig:production}Neutrino yields for the SNS mercury and
  ESS tungsten targets, as a function of simulation package
  adopted. The first column shows predictions from a dedicated
  calculation \cite{burman,report}, validated against neutrino cross
  section measurements \cite{louis}. For three physics lists in common
  with \cite{bench1}, arrows attempt a correction based on the
  HARP-CDP $\pi^{+}$ production cross-section for 2.2 GeV protons on
  Ta. The horizontal red line marks the value used for the CE$\nu$NS
  measurement at the SNS \cite{science}, whereas the black line shows
  the value adopted here for the ESS. }
\end{figure}

{Nevertheless in this work we perform our own simulations
as an additional test of the neutrino production rates at the ESS.
In particular we have performed simulations with MCNPX \cite{mcnpx},
GEANT4 \cite{g4}, and FLUKA \cite{fluka} shown in Fig.~\ref{fig:production}.} 
While MCNPX was used for the baseline design and
target optimization of the ESS \cite{essdesign,essopt}, recent
comparisons with GEANT4 qualify the latter as a comparable tool, at
least from the point of view of neutron production and transport
\cite{mvsg}. Fig.\ \ref{fig:production} shows a typically fair
agreement between all codes and the calculated neutrino production in
\cite{burman,report}, for a proton energy of 0.94 GeV. This agreement
is preserved at 2 GeV for most MCNPX 2.7.0 intranuclear cascade and
evaporation model combinations, but not for the GEANT4 options
tested. A large dispersion in GEANT4 pion yield predictions has been
noticed before, during benchmarks using the most recent
hadroproduction measurements by the HARP \cite{harp} and HARP-CDP
\cite{harpcdp} collaborations: experimental pion production
cross-sections for 2.2 GeV protons striking Ta and Pb targets are
over- or under-estimated by different GEANT4 physics lists
\cite{bench1,bench2,bench3}, following a pattern similar to that in
Fig.\ \ref{fig:production}. In addition to this, a comparison with
HARP-CDP data \cite{rec1} indicates that $\pi^{+}$ production by 0.8
GeV protons is slightly underestimated by the LAHET parametrization in
\cite{burman,report}.

%\begin{widetext}
%\end{widetext}

Based on these considerations, we tentatively adopt a yield of 0.3
neutrinos of each flavor ($\nu_{\mu},\bar{\nu}_{\mu},\nu_{e}$) per
proton for a ESS operating at 2 GeV, a considerable improvement
{of a factor 3.75 over
the corresponding figure of 0.08 at the SNS \cite{science}
(but a bit  more conservative than the 4.4 increase factor
found in Refs.\cite{burman,report})}.
In combination with the increased ESS proton current mentioned above,
this results in an expected 8.5$\times 10^{22}$ neutrinos per flavor
per year, an order of magnitude higher than the equivalent of
9.2$\times 10^{21}$ from a reference 1 MW, 0.94 GeV SNS
\cite{jason}. On the subject of neutrino flux verification, a proposal
to use a small (1 m$^{3}$ D$_{2}$O) charged-current detector to
independently measure the neutrino yield of the SNS
\cite{jason,yurinue,d2o} may be worth replicating at the ESS.

A second difference between SNS and ESS is in their proton beam pulse
timing: 60 Hz of 1 $\mu$s-wide POT spills at the SNS, vs.\ 14 Hz of
2.8 ms spills at the ESS (Fig.\ \ref{fig:ess}). Naively, beam timing
at the SNS would seem much more favorable for steady-state background
reduction in CE$\nu$NS detectors \cite{scholberg}. The SNS duty factor
is $6\times 10^{-4}$, accounting for 10 $\mu$s after POT pulses to
encompass the delayed $\overline{\nu}_{\mu},\nu_{e}$ neutrino emission
from muon decay \cite{science}: this duty factor becomes $4\times
10^{-2}$ at the ESS. However, steady-state backgrounds can be
accurately characterized during the long anti-coincident periods
between beam spills, to then be subtracted from POT-coincident windows
that in addition contain the CE$\nu$NS signal. In the absence of
beam-related backgrounds, this results in a residual departing from
zero proportionally to the CE$\nu$NS signal, with statistical error
bars partly defined by the steady-state background level achieved
\cite{science}. The signal-to-background figure of merit is then seen
to be slightly favorable for the ESS (the square root of the ratio of
the ESS and SNS duty factors is 10\% smaller than the factor of nine
increase in CE$\nu$NS signal rate from a larger ESS neutrino flux).

Regardless of this minor advantage, the main attraction of the ESS is
in the accumulation of CE$\nu$NS signal statistics ten times faster
than at the SNS. An example of the relevance of this aspect can be
found in the sluggish signal growth rate for the 14.6 kg CsI[Na]
detector at the SNS: some $\sim$300 events in four calendar years, for
a heavy-nuclei target with a large CE$\nu$NS cross-section. In
contrast to this, the similarly compact detectors presented in
Sec.~\ref{sec:detectors}
aim to register up to thousands of events per year at the
ESS (Table~\ref{tab:detectors}), through a combination of increased
neutrino flux, and improved detector performance.

Related to the timing consideration above, at the ESS there will be no
temporal separation possible between neutrino flavors, with prompt
$\nu_{\mu}$ and delayed $\bar{\nu}_{\mu},\nu_{e}$ becoming
indistinguishable due to the very long (2.8 ms) beam spills. At the
ESS, a partial discrimination is nevertheless possible using recoil
energy spectrum, which reduces the impact of this limitation for most
NP searches, as will be discussed Sec.~\ref{sec:nsi}. There are however a few exceptions where a distinction between prompt and delayed
signals is mandatory, such as the search schemes described in
\cite{carlos,dm3}. As shown by COHERENT \cite{science} an opportunity
exists at the SNS to (partially) isolate prompt from delayed neutrino
components via timing, but only for detectors sited in a small area of
the ``neutrino alley" \cite{science}, of roughly 10$\times$1
m$^{2}$. This limitation is due to a large gradient in prompt neutron
background measured along the length of the alley (five orders of
magnitude over 25 m \cite{bjorn,rex,rex2}), and safety restrictions on
corridor width encumbrance. The end of the SNS alley farthest from
target (28 m) offers additional space for detectors requiring
ancillary equipment such as cryogenic and purifier stages for liquid
noble targets. However, the prompt neutron background from the nearby
beam line makes $\nu_{\mu}$ detection entirely unfeasible at this
location \cite{rex,rex2}, further reducing the available SNS neutrino
flux by one third.

Still on the subject of space availability, a strong underlying
assumption in this work is that a ESS location will be found that is
similar to the SNS neutrino alley, in its optimal proximity to target,
shielding against prompt neutrons, and absence of interference with
neutron activities. Specifically, an instantaneous (i.e.,
POT-coincident) flux of less than $\sim2\times10^{-3}$
neutrons/cm$^{2}$ s ($E_{n}>$1 MeV) will be required at the ESS to
achieve a negligible beam-coincident neutron background level, similar
to that in \cite{science}. The lessons learned in this respect from
SNS experimentation (neutron leakage from beam line, influence of
materials and voids in the line of sight to the target monolith,
beneficial effect of basement overburden against skyglow) can be used
to minimize the scouting for such locations. Initial investigations of
the availability of ESS sites viable for CE$\nu$NS studies indicate
that 22 m$^{2}$ of unallocated space 15-23 m from target exists, with
5 meters of steel or iron and a minimum of 6 meters of
magnetite-loaded heavy concrete in the line of sight to target. The
density of this concrete is twice that of the compacted limestone
gravel between target monolith and neutrino alley at the
SNS. Dedicated neutron background measurements and simulations will be
necessary to confirm the suitability of this and other ESS locations
for the activities described next.

\section{Detector Technologies}
\label{sec:detectors}

In this section we present a number of detector technologies
sensitive
to low-energy ($\lesssim$1 keV$_{nr}$) nuclear recoils, aligned to
profit from the increased neutrino flux at the ESS.
Some possess characteristics able to maximize the
physics reach to certain physics scenarios, and
in combination they can further boost the
attainable sensitivity, as we illustrate in  Sec.\ref{sec:pheno}.
The common origin of these techniques is in the fields of dark matter
detection, and double-beta decay searches.
A best-effort in background abatement
already made for these detectors can be reinvested for CE$\nu$NS at
the ESS. For convenience, the characteristics of the detectors considered,
and expected signal and background rates assumed in the sensitivity studies
presented in Sec.~\ref{sec:pheno} are summarized in Table ~\ref{tab:detectors}.

\subsection{Cryogenic (77 K) undoped CsI scintillator array}

\begin{figure}[!htbp]
\centering  
\includegraphics[width=1\columnwidth]{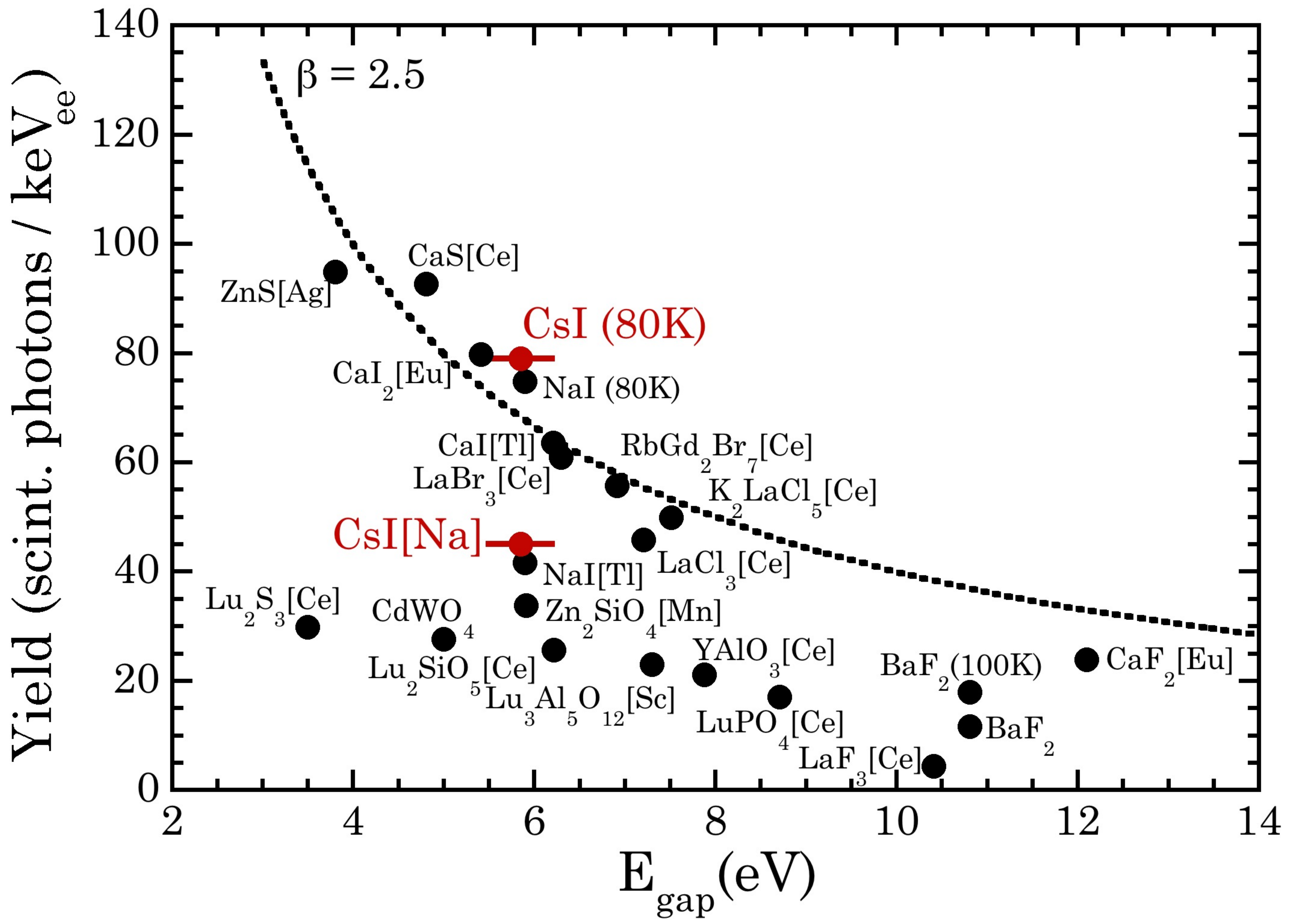}
\caption{\label{fig:ly} Light yield of scintillators and phosphors as a function of bandgap, adapted from \cite{maximumLY}. A dotted line indicates a model-dependent maximum light yield  \cite{maximumLY}. Red dots mark our measurements for cryogenic CsI, in  agreement with \cite{liu,zhang}, and for room temperature CsI[Na] \cite{ournim}. Yields of up to 125 ph/keV$_{ee}$ have been claimed for other CsI stock \cite{mos1,mos2}. A CsI  bandgap range is indicated by horizontal error bars \cite{bandgap1}.}
\end{figure}

Pure CsI operated at liquid nitrogen temperature exhibits a light yield in the range  80-125 photons per keV \cite{amsler,mos1,mos2,nadeau,clark,liu,woody,zhang,mik}. This is at the maximum theoretical efficiency in the conversion of energy deposition to scintillation, with almost all electron-holes created in the crystal recombining radiatively (Fig.\ \ref{fig:ly}, \cite{csiisspecial,maximumLY}). This behavior is remarkable, as the number of information carriers generated per unit deposited energy is just a factor of three smaller than for Ge and Si semiconductor detectors. Provided that a good quantum efficiency (QE) in the light sensor is achieved, this can facilitate the detection of low-energy signals, with optimal resolution. The potential of cryogenic CsI for neutrino and dark matter detection has attracted attention \cite{nadeau,clark,liu,zhang,angloher}. This material further improves on the virtues of CsI[Na] for CE$\nu$NS \cite{ournim} , by increasing its light output by a factor of $\gtrsim$ 2 (Fig.\ \ref{fig:ly}). 

%\begin{widetext}
%\end{widetext}

%\begin{widetext}
\begin{figure}[!htbp]
\centering  
\includegraphics[width=1\columnwidth]{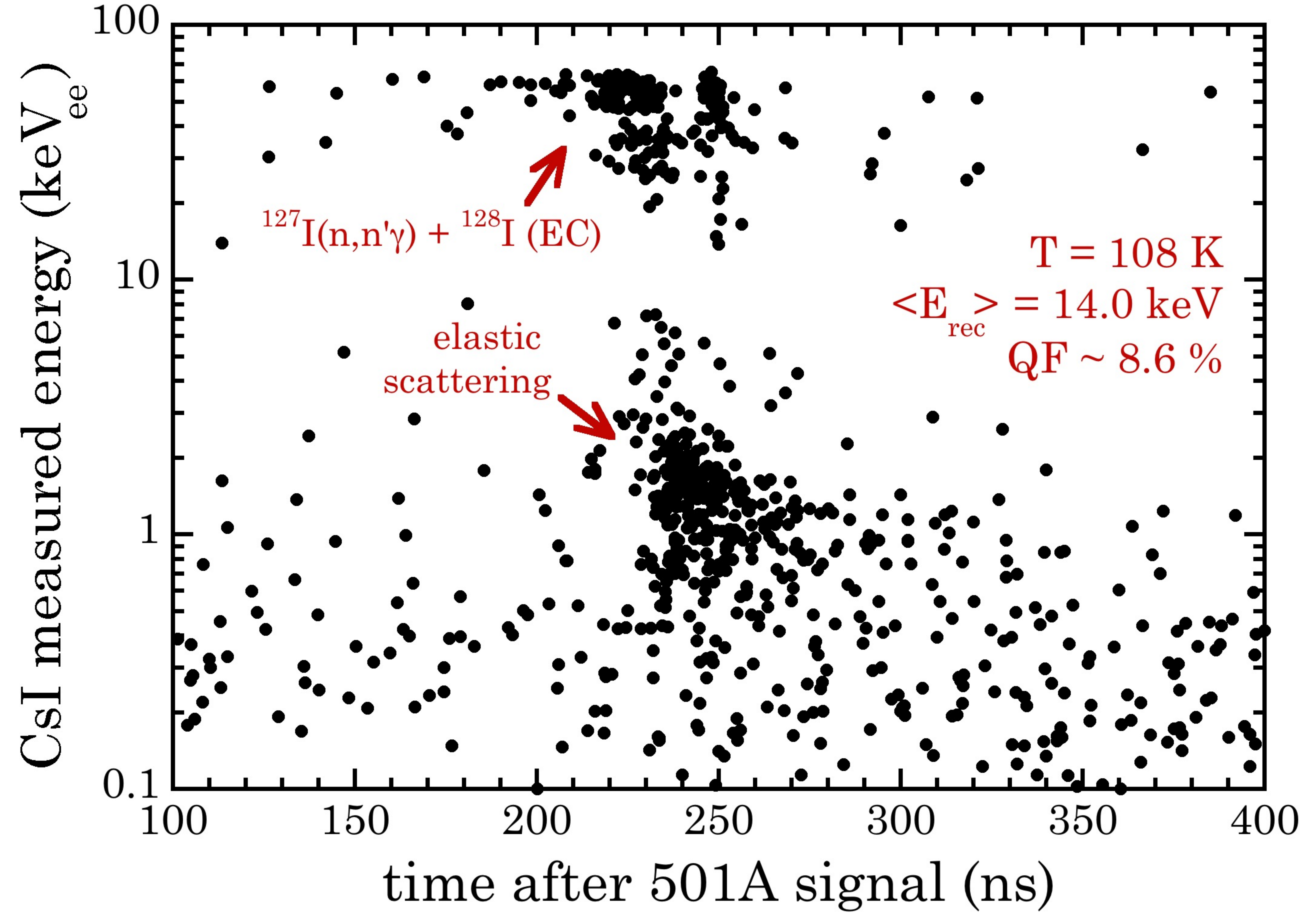}
\caption{\label{fig:QF} First measurement of nuclear recoil quenching factor (QF) for cryogenic CsI. The figure displays the measured electron-equivalent energy (keV$_{ee}$) deposited by 14 keV Cs and I nuclear recoils induced by 2.2 MeV neutron scattering. The reader is referred to a similar Fig.\ 2 in \cite{csiqf}, for more details on the methodology employed in QF determination. }
\end{figure}
%\end{widetext}

%\begin{widetext}
\begin{figure}[!htbp]
\centering
  \includegraphics[width=1\columnwidth]{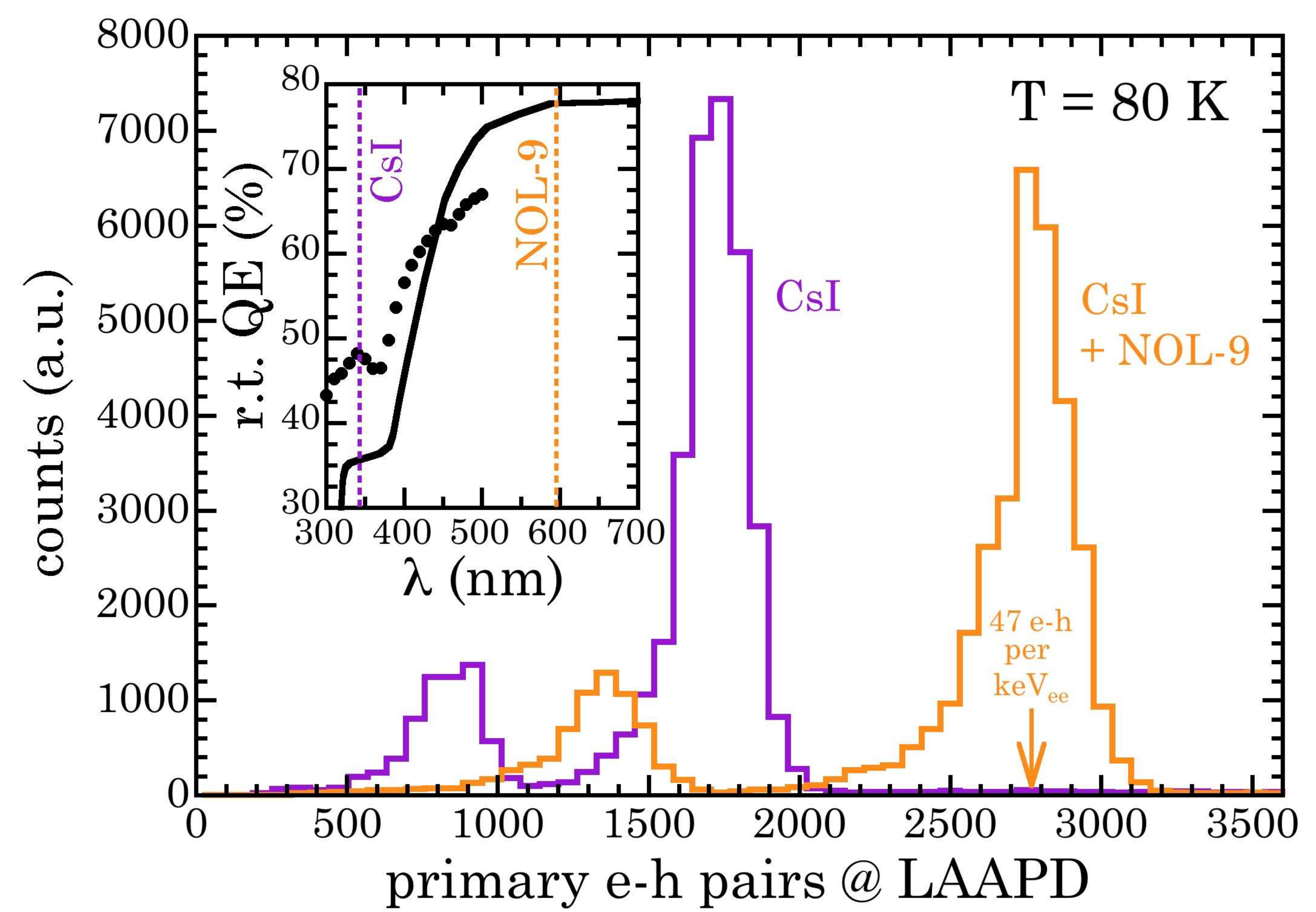}
\caption{\label{fig:nol9}  Response of a 3.2 cm$^{3}$ cryogenic CsI
  crystal to 59.5 keV $^{241}$Am gammas, seen by a $1.3\times1.3$
  cm$^{2}$ LAAPD \cite{rmd}, with and without a NOL-9 wavelength shifter
  plate \cite{belle,jin}. Escape peaks appear at 55.5 keV and 29.7
  keV. LAAPD gain and noise (G=1,060, 4 eV rms) were monitored via
  concurrent silicon surface irradiation with $^{55}$Fe x-rays
  \cite{mos1,mos2}. Plasma effects \cite{saturation} and QE reduction
  with temperature \cite{yang} were considered. {\it Inset:} available
  room-temperature (r.\ t.\!) QE data from the LAAPD manufacturer
  (dots), and for a generic Si APD (line) \cite{pmthandbook}, together
  with wavelengths of CsI emission (80 K, \cite{woody,amsler}) and of
  NOL-9 r.\! t.\ luminescence \cite{lumin}. The increase in photon
  detection is as expected from an efficient wavelength shift. A 1 keV
  nuclear recoil with QF = 10\% will generate a 4.7 primary e-h
  signal, above LAAPD threshold (see text).}
\end{figure}
%\end{widetext}

In order to establish the feasibility of using cryogenic CsI at the
ESS, two cryostats have been developed at the University of
Chicago. One is dedicated to large-area avalanche photodiode
(LAAPD)~\cite{mos1,mos2,amsler,yang,exo} readout of CsI crystals at 80
K. The second is reserved for Hamamatsu R8520-506 \cite{cryopmt}
photomultiplier (PMT) use at 108 K. This ongoing R\&D is of crucial
importance to establish the method: for instance, the quenching factor
(QF) for low-energy nuclear recoils (NRs) in pure cryogenic CsI has
not been measured before. Our preliminary findings at 108 K
(Fig.\ \ref{fig:QF}) indicate that it is of O(10)\%, comparable to
that for CsI[Na] at room temperature \cite{csiqf}. An anomalous
increase of the QF for alphas in CsI at low temperature
\cite{nadeau,clark}, and the possible presence of coadjutant
low-energy processes such as the Migdal effect \cite{migdal}, are a
reminder of the need for a full QF characterization. We temporarily
adopt an energy-independent value of 10\% (Table~\ref{tab:detectors}).

As part of the upgrade of the Belle-II CP-violation experiment, novel
wavelength shifters (nanostructured organosilicon luminophores, NOL
\cite{lumin,nol,nol2,nol3,nol4,nol5}) have been used to reach a quantum efficiency QE
$\gtrsim$ 80\% during avalanche photodiode (APD) monitoring of
room-temperature CsI \cite{belle,jin}. Fig.\ \ref{fig:nol9} displays a
satisfactory first demonstration of NOL performance at cryogenic
temperature, using our LAAPD cryostat. The light-detection QE obtained
following wavelength shifting of the 340 nm cryogenic CsI emissions
\cite{amsler,woody} to 590 nm is a factor of three larger than
possible with existing cryogenic PMT photocathodes \cite{cryopmt}. We
observe an excellent long-term stability in the performance
(electronic noise, internal gain, light yield) of the CsI+NOL+LAAPD
combination at 80 K.

\begin{figure}[!htbp]
\centering  
\includegraphics[width=1.01\columnwidth]{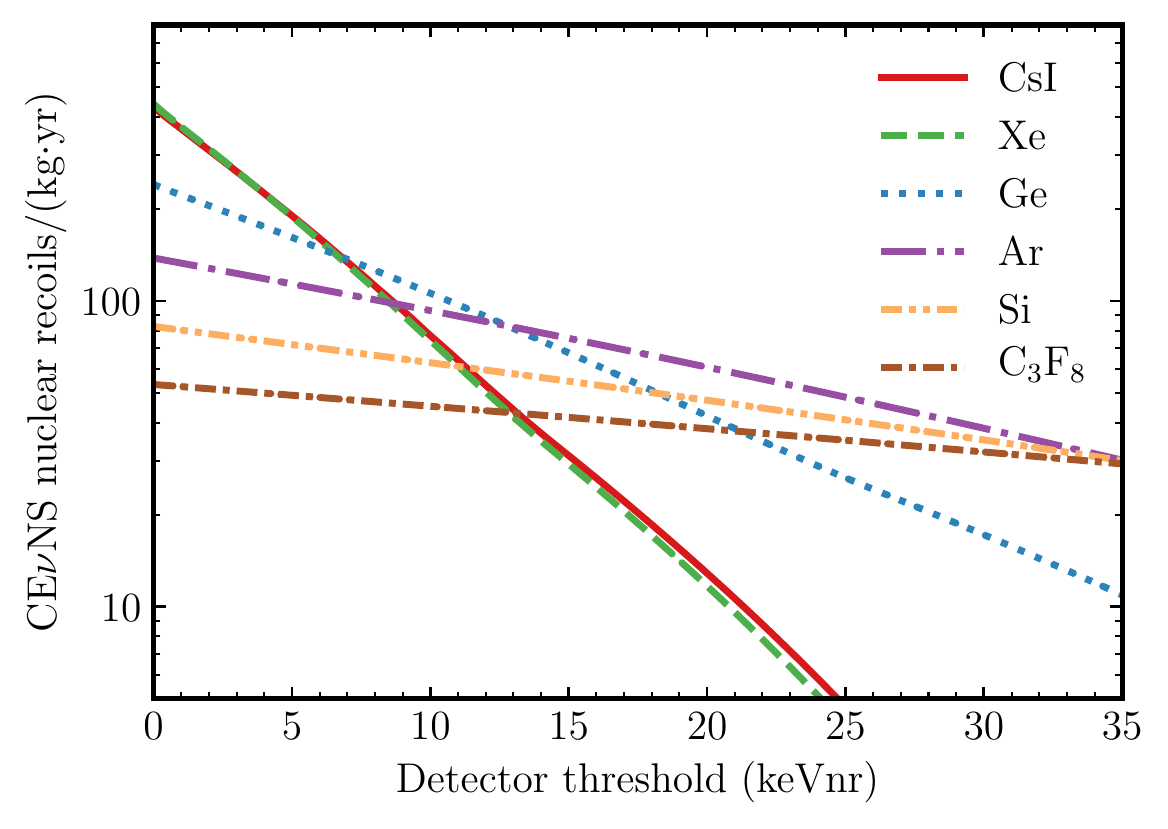}
\caption{\label{fig:rates} Expected integrated CE$\nu$NS rate above
  nuclear recoil threshold, 20 m away from the ESS target, for all
  detector materials considered in this work. All technologies
  considered for use at the ESS have thresholds at or below 1
  keV$_{nr}$.}
\end{figure}

LAAPDs with surface area 45 cm$^{2}$, when operated at 77 K, exhibit
stable internal gains in excess of 1,000, and a reduced leakage
current leading to a light-detection threshold of approximately four
photons \cite{farrell}, i.e., four primary electron-hole (e-h) pairs
at the LAAPD (Fig.\ \ref{fig:nol9}) prior to avalanche
amplification. The demonstrated combination of a high light yield
(Fig.\ \ref{fig:ly}), optimal QE for its detection
(Fig.\ \ref{fig:nol9}), few-photon LAAPD threshold \cite{farrell}, and
a quenching factor of order 10\% (Fig.\ \ref{fig:QF}), adds up to a
predicted sensitivity to $\sim$1 keV nuclear recoils in this hybrid
cryogenic detector. This can be contrasted with the $\sim$10 keV
obtained with super-bialkali PMT readout of room temperature CsI[Na]
during the first CE$\nu$NS measurement
\cite{science,bjorn}. Furthermore, CsI[Na] NR signal acceptance was
limited there to 65\%, as a result of cuts to reject Cherenkov light
emission in the PMT glass envelope. This is a dominant source of
low-energy background that is absent from LAAPDs. Taking the effect of
signal acceptance into account, Fig.\ \ref{fig:rates} shows that an
increase by a factor of eight in CE$\nu$NS signal rate per unit CsI
mass can be obtained from this alternative cryogenic approach. The
steady-state background adopted in Table~\ref{tab:detectors} is that
achieved for CsI[Na] at the SNS \cite{science,bjorn}, with the
estimated Cherenkov contribution subtracted.

%\begin{widetext}
%\end{widetext}

The presently envisioned design of a ESS cryogenic CsI detector is a
small array of four crystals of individual dimensions
$5\times5\times50$ cm$^{3}$, each read out by two 25 cm$^{2}$ LAAPDs,
for a total mass of 22.5 kg.  This compact detector is expected to
provide approximately 8,000 CE$\nu$NS events per year at the ESS
(Fig.\ \ref{fig:rates}), a signal throughput two orders of magnitude
faster than during the first CE$\nu$NS detection \cite{science}.
Three hundred CsI crystals of these dimensions are in storage at the
University of Chicago \cite{ktev}, left from the KTeV experiment,
allowing for an eventual detector mass upgrade.

\subsection{Low-background CCD arrays with single-electron threshold}

High resistivity, $\approx$mm-thick silicon Charge Coupled Devices
(CCDs) have been recently demonstrated as an effective detector
technology for the search of rare events from dark matter
\cite{damicsnolab, damicelectron,sensei} and neutrino \cite{connie}
interactions. Efforts are ongoing for the construction of
DAMIC-M~\cite{damicm}, a kg-size detector to be installed at the
Laboratoire Souterrain de Modane (LSM) in France. To achieve a leading
dark matter sensitivity, DAMIC-M will require background rates at the
LSM of 0.1 counts per keV$_{ee}$-kg-day (ckkd), and a threshold of two
ionized electrons (corresponding to a few eV in ionization
energy). This ultra-low energy threshold and background rate can be
leveraged for CE$\nu$NS detection. As discussed in
Sec.~\ref{sec:pheno}, the sensitivity to a finite neutrino magnetic
moment via CE$\nu$NS becomes maximal at the lowest recoil energies. A
detector with characteristics like DAMIC-M in threshold and energy
resolution would be ideal for this CE$\nu$NS application. In the
following we will use DAMIC-M as an example for a kg-size CCD-based
silicon detector to be installed at the ESS. DAMIC-M will be nearing
completion of its exposure at the LSM by the time the ESS reaches its
full 5 MW power.

\begin{figure}[!htbp]
\centering  
\includegraphics[width=1.08\columnwidth]{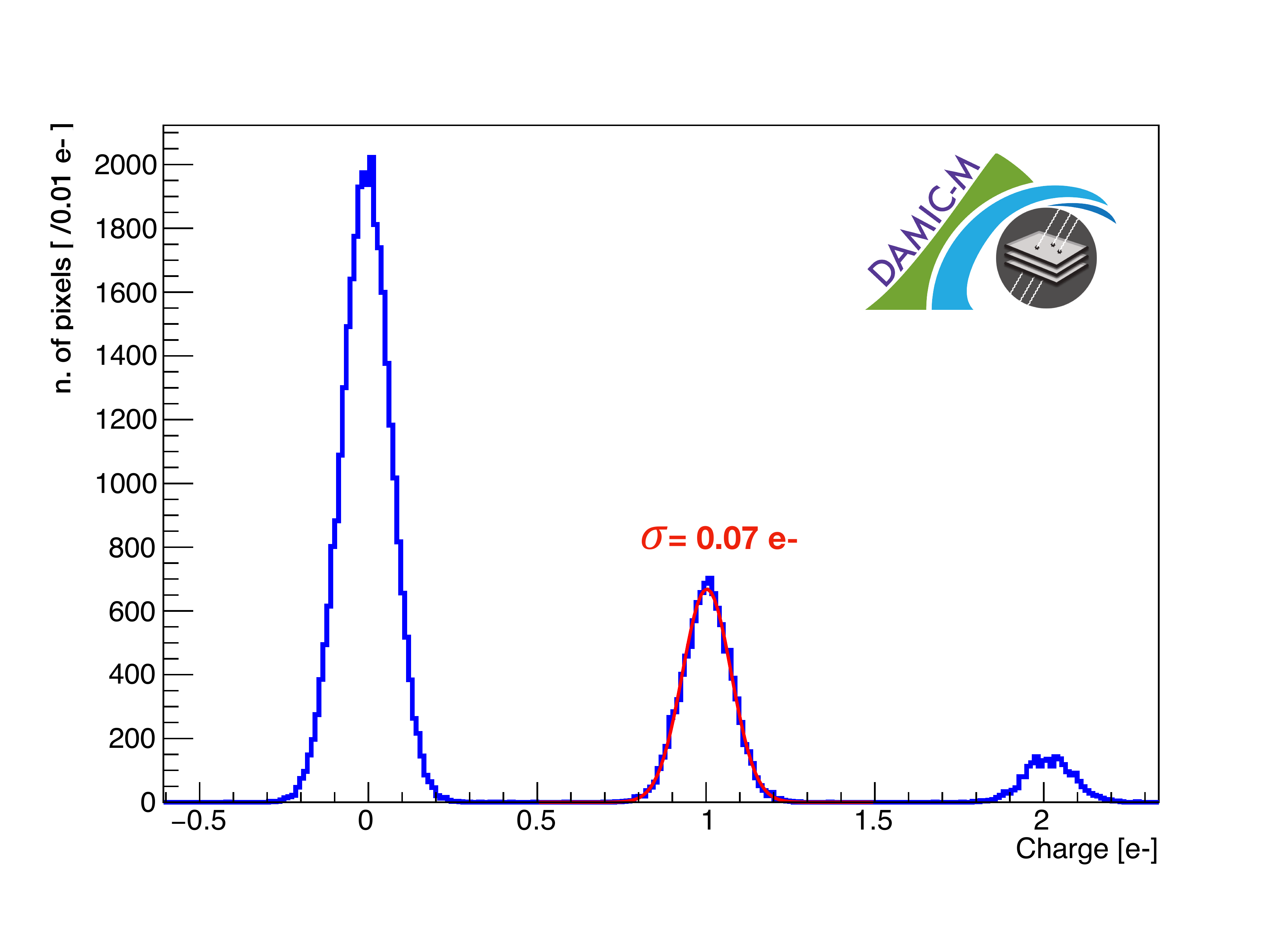}
\vglue -0.8cm
\caption{\label{fig:damic} (From \cite{priviterataup}) First
  demonstration of single-electron sensitivity in DAMIC-M CCDs, using
  a Skipper readout. An average ionization energy deposition of 3.77
  eV is necessary for the creation of a 1 e$^{-}$ signal in silicon
  \cite{esi}. }
\end{figure}

The beam timing characteristics of the ESS involve a data-acquisition
approach different from that of a dark matter search. For a typical
single pixel read-out time of 10~$\mu$s the total read-out time for a
Mpixel device would be several seconds, too slow for the ESS
beam-spill frequency. We thus plan to perform "hardware-binning"
(namely to sum the signal of several pixels before read-out), a
technique which improves read-out speed and read-out noise with only
minimal loss of event information. Hardware binning has been
successfully employed by DAMIC@SNOLAB~\cite{damicsnolab}. The number
of binned pixels that can be readout given the ESS beam-spill
frequency of 14 Hz will depend on the number of amplifiers integrated
in the CCD, with about 7000 binned pixels per amplifier per beam
spill. Taking the current design of a DAMIC-M CCD as example (6k
$\times$ 6k pixels, each 15 $\mu$m $\times~ 15~\mu$m, readout by four
amplifiers), 14,000 binned pixels can be readout per spill when
accounting for identical 2.8 ms POT-coincident and -anticoincident
exposures per beam spill (the second used to characterize steady-state
backgrounds). This can be achieved by a 50$\times$50 binning
(corresponding to a 750 $\mu$m$ ~\times 750~\mu$m sensitive area) or
any other suitable combination of rows and columns summing a total of
2500 pixels. No hardware modifications are required for this purpose.
Future CCDs could be better optimized for ESS timing by incorporating
more amplifiers for faster read-out.

Instrumented with a Skipper readout \cite{javier}, DAMIC-M will
feature single-electron sensitivity. DAMIC-M has recently demonstrated
such sensitivity using 1k $\times$ 6k CCDs to achieve a resolution of
less than 0.1 electrons \cite{priviterataup}, as shown in
Fig.\ \ref{fig:damic}. This is accomplished by using a novel approach
to data acquisition, wherein the charge in a single pixel is measured
non-destructively multiple (N) times resulting in a reduction of the
overall read-out noise by a factor of $\sqrt{N}$. While this certainly
could not be implemented for every pixel at the ESS, an adaptive
read-out mode wherein a pixel is measured multiple times only if it
meets a set of selection criteria may provide improved energy
threshold with minimal loss of exposure.

As mentioned, DAMIC-M is designed to reach a 0.1 ckkd internal
radioactive background, far lower than other solid-state detectors of
the same scale.  The steady-state background for a silicon CCD
detector at the ESS is therefore expected to be cosmic-ray associated,
and dominated by muon-induced neutrons in lead shielding, which can be
rejected using an active veto. Instead of discarding individual
events, as is done for faster detectors, entire veto-coincident
individual CCD exposures must be rejected. Accounting for a 100 Hz
muon veto trigger rate under minimal overburden, and for the ESS
beam-timing characteristics, we estimate that this can be achieved
with just $\sim$28 \% dead time, while rejecting $>$99.9 \% of
muon-induced backgrounds. These expectations are solidly based on the
germanium PPC experimentation at shallow depth described below. In the
absence of a measurement of background level at a shallow site, we
presently assume that the background rates at the ESS will be a factor
of ten higher than those expected by DAMIC at the LSM, and thus adopt
a background level of 1 ckkd in the sensitivity estimates of
Sec.~\ref{sec:pheno}. The 15 ckkd at 0.18 keV$_{ee}$ obtained in a
shallow site with less-radioclean PPCs (Fig.\ \ref{fig:ppc}) is a
realistic upper bound.

The quenching factor for DAMIC silicon CCDs has been measured down to
0.7 keV$_{nr}$ \cite{alvaro}, using a photoneutron source
\cite{photoneutron}.  A new measurement down to a considerably lower
energy is planned using the same technique, while profiting from the
recent progress in single-electron sensitivity. As emphasized in
\cite{csiqf}, a precise knowledge of this quenching factor will be
necessary to fully exploit the sensitivity of CE$\nu$NS detectors to
NP.  The incipient but rapid development of new germanium CDDs
\cite{geccd} is worth mentioning as a possible future upgrade of this
technique, as those can provide a higher CE$\nu$NS signal rate
(Fig.\ \ref{fig:rates}), and a more favorable quenching factor.

\subsection{High-pressure gaseous xenon chambers}

\begin{figure}[!htbp]
\includegraphics[width=1.02\columnwidth]{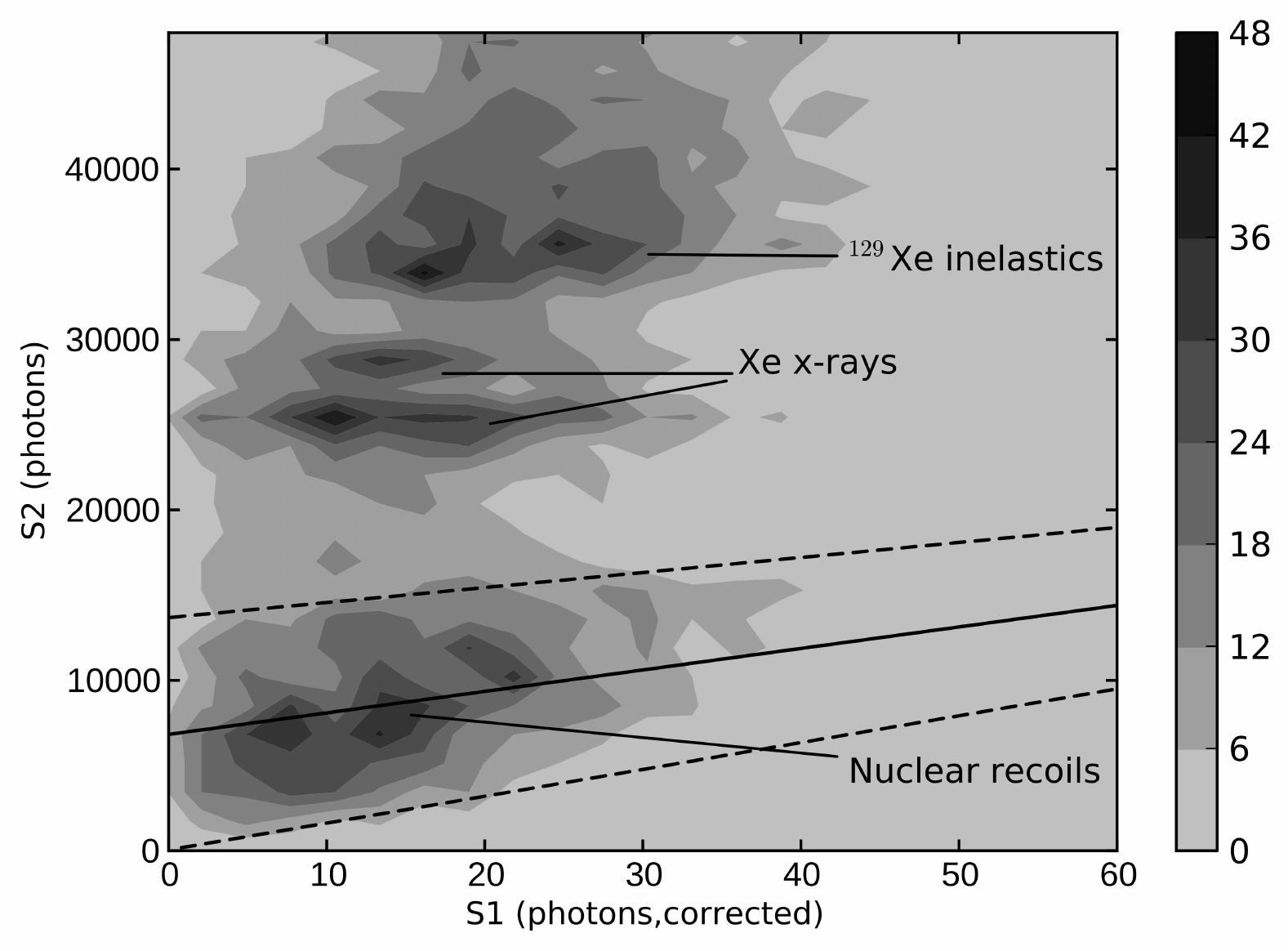}  
\includegraphics[width=1\columnwidth]{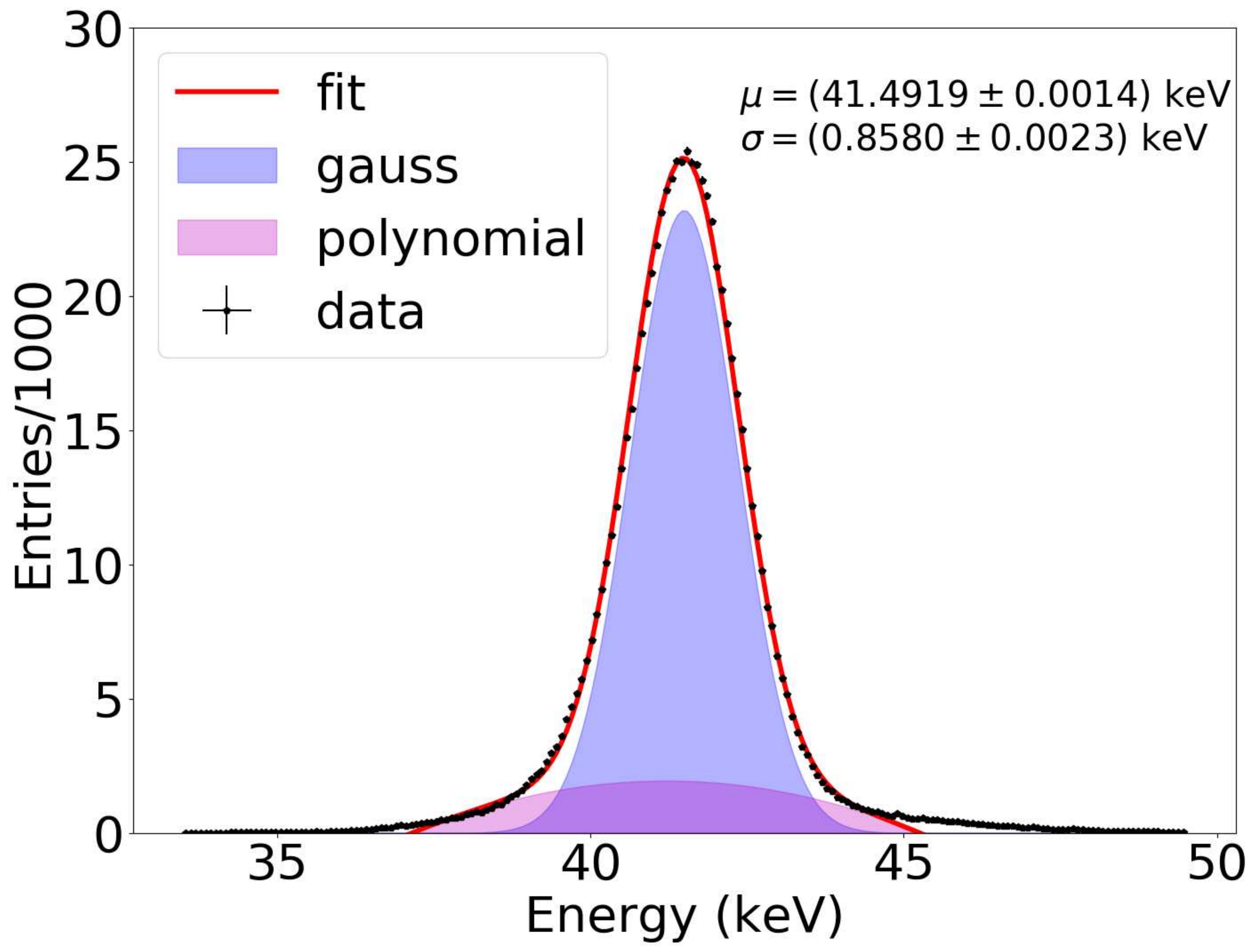}
\caption{\label{fig:xe_res}Left: Nuclear recoil identification in a
  gaseous xenon detector by using S1/S2 discrimination
  \cite{gas_xenon_nuclear}. Right: Example of energy resolution
  obtained for 41.5 keV gammas \cite{kr_res}. The very small Fano
  factor in xenon gas, two orders of magnitude lower than for liquid
  xenon, allows for excellent energy resolution even in large-scale
  detectors \cite{next100}.}
\end{figure}

High-pressure gaseous detectors such as NEXT \cite{next}, sensitive to
both primary scintillation (S1) and electroluminescent amplification
of charge ionization (S2), provide excellent background
discrimination, and optimal energy resolution
(Fig. \ref{fig:xe_res}). This technology is not impacted by charge
trapping and delayed release at the liquid-gas interface of dual-phase
liquid noble gas detectors \cite{sorensen,trapping}. This process
limits dual-phase detector sensitivity to low-energy signals at
shallow depth sites dominated by frequent large-energy depositions by
cosmic rays \cite{xesurf}, making NEXT-like detectors more suitable
for operation at a location with negligible overburden, like the ESS.

Furthermore, in principle the ability to trigger data acquisition
using a POT-coincident logic signal can eliminate the need to detect
primary scintillation light at the ESS, resulting in a reduced energy
threshold. By using electroluminescence amplification, signals as low
as 1-2 ionized electrons can be detected. This reduces the expected
energy threshold to less than 0.2 keV$_{ee}$. In addition to this,
gaseous xenon detectors feature a few-percent energy resolution at low
energy (Fig. \ref{fig:xe_res}), only surpassed by semiconductor
detectors. Relaxing the need to detect primary scintillation would
result in the loss of background rejection ability. However,
background rejection has been shown not to be a requirement for
CE$\nu$NS detection, for sufficiently radioclean detectors such as the
CsI[Na] crystal in \cite{science}, or the germanium PPCs discussed
below (Fig.\ \ref{fig:ppc}).

Dedicated studies of the response of gaseous detectors to few-keV
nuclear recoils will be necessary to reduce the present uncertainty on
parameters such as the quenching factor. Such measurements are planned
using a photoneutron source \cite{photoneutron} in a 1-kg scale
NEXT-like detector. In the interim we use a QF=20\%, similar to that
adopted for S2 generation in \cite{gas_xenon_nuclear}.

One interesting possibility for this detector design is the ability to
use different noble gas targets within the same setup. This will allow
to compare data taken with xenon, krypton, argon, neon, and even
helium. At the time of this writing, a large-volume gaseous detector
is already searching for neutrinoless double-beta decay, using high
pressure xenon gas: the NEXT-White detector \cite{new}. This device,
with internal dimensions of 0.57 m diameter $\times$ 0.72 m length can
hold up to 20 kgs of Xe at an operating pressure of 20 bar. Once
replaced by the planned upgrade to NEXT-100 in 2020 \cite{next100},
NEXT-White can be easily adapted to low-energy searches, and used for
CE$\nu$NS studies at the ESS. In the present absence of a dedicated
background measurement at shallow depth, we adopt in
Table~\ref{tab:detectors} a level similar to that for CsI.

\subsection{Low-threshold, multi-kg p-type point contact germanium detectors}

P-type point contact germanium detectors (PPCs, \cite{ppc}) provide a
unique combination of detector mass, ultra-low energy threshold, and
background rejection capabilities. As such they have found multiple
applications in neutrino (Majorana \cite{majorana}, GERDA
\cite{gerda}, TEXONO \cite{texono}, CONUS \cite{conus}) and dark
matter searches (CoGeNT \cite{cogent}, CDEX \cite{cdex}).  A
continuous improvement in PPC mass and energy threshold
\cite{gabriela} has resulted in large devices approaching 0.1
keV$_{ee}$ sensitivity. Inverted coaxial PPCs \cite{radford} allow for
single multi-kg crystals.

\begin{figure}[!htbp]
\centering
\includegraphics[width=1\columnwidth]{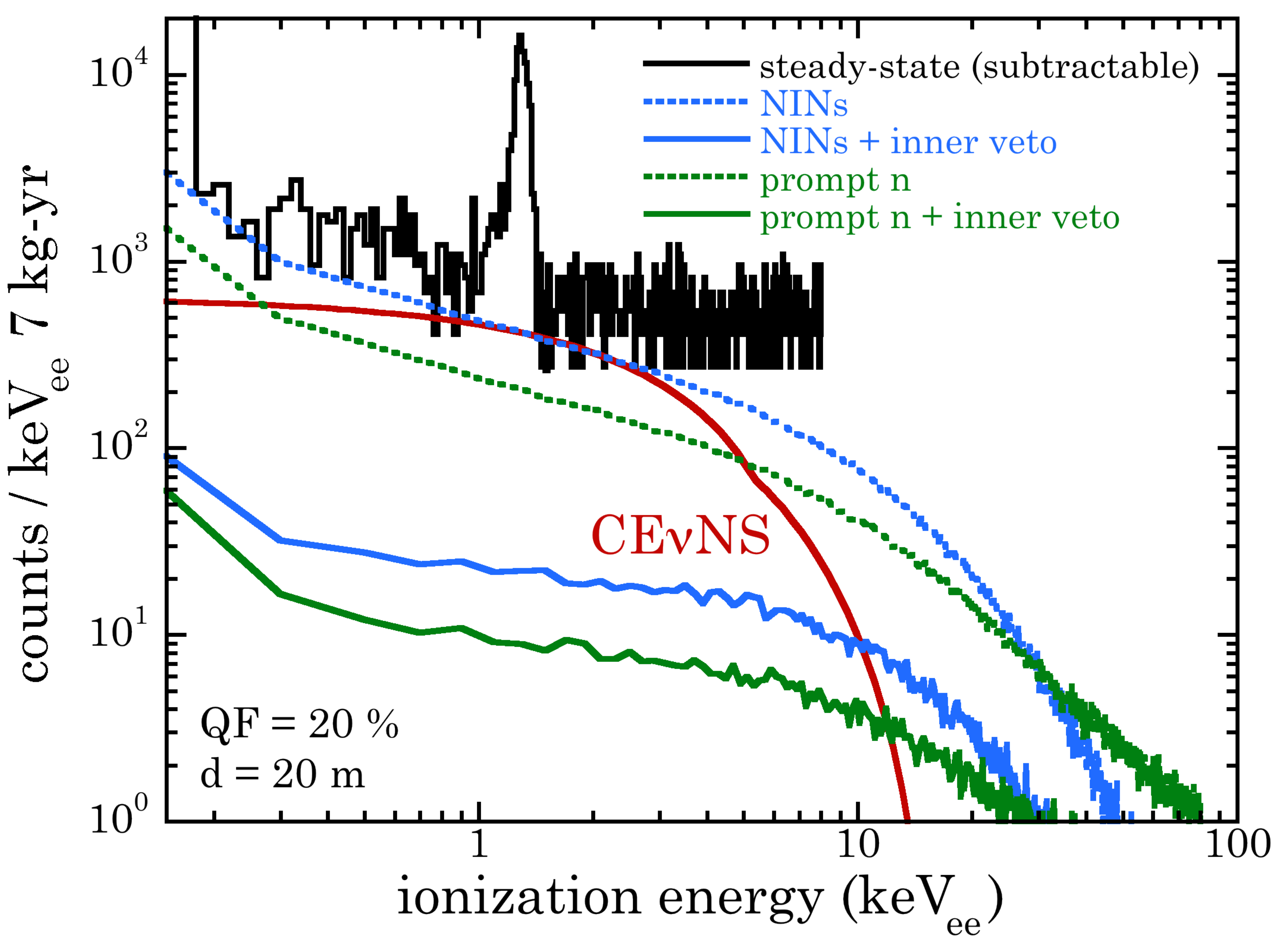}
\caption{\label{fig:ppc} Comparison of expected CE$\nu$NS signal in Ge
  PPCs at the ESS to backgrounds. The beam-associated prompt neutron
  flux and NIN production rate are taken to be $\times$10 those at the
  SNS \cite{science}, with same spectral hardness, in the simulations
  shown. The effect of the inner veto uses as input its measured light
  collection efficiency, and a neutron response for plastic
  scintillator as in \cite{plastic}. The subtractable steady-state
  background achieved at 6 \protect\frenchspacing{m.w.e.} with the PPC
  described in the text is shown, applying the ESS duty factor for a
  direct comparison to CE$\nu$NS. As a reference, the CE$\nu$NS signal
  to steady-state background ratio for CsI[Na] in \cite{science} was a
  less favorable $\sim$1/4.}
\end{figure}

A recently completed 2.95 kg PPC features a stable 0.18 keV$_{ee}$
threshold, and a 15 ckkd background at threshold (Fig.\ \ref{fig:ppc})
during operation in a shallow overburden site at the University of
Chicago (6 \frenchspacing{m.w.e.}). This is similar to what can be
expected in a ESS basement location (the SNS neutrino alley provides 8
\frenchspacing{m.w.e.}). This background is achieved via a new
shielding design that includes a double active veto. Its innermost
plastic scintillator layer surrounds the PPC. It is able to tag the
neutrino-induced neutron (NIN) background from lead shielding
\cite{science}, and beam-related prompt neutrons, with a high
efficiency (Fig.\ \ref{fig:ppc}). The shield design is intentionally
compact at 60 cm x 60 cm x 150 cm. Use of a cryocooler provides
unattended operation without access to liquid nitrogen for periods
$>$1 year. Special measures were taken in the internal detector design
and FPGA-based data-acquisition to obtain an absence of measurable
cryocooler-induced microphonic noise. The device is presently unique
in its combination of large mass and low energy threshold. However, a
further reduction in PPC leakage current is planned, aiming to push
the threshold down to $\sim$0.12 keV$_{ee}$.  An effort towards the
characterization of the sub-keV quenching factor in germanium is also
underway \cite{csiqf}. For the purposes of ESS sensitivity
calculations, we adopt a germanium mass of 7 kg (achievable with two
PPCs), the already demonstrated background of Fig.\ \ref{fig:ppc}, the
upgraded threshold, and a quenching factor of 20\% \cite{geqf} (see
Table~\ref{tab:detectors}).

\subsection{Moderately superheated liquids}

Moderately superheated bubble chambers like those used for the PICO
dark matter search \cite{chambers,coupp,pico} provide a dramatic
insensitivity to electron recoil (ER) backgrounds, the best of any
nuclear recoil detector \cite{baxter}. A recent development in this
area are scintillating bubble chambers \cite{eric,eric2}. The
additional information channel provided by light detection facilitates
sub-keV nuclear recoil thresholds, while preserving ER
insensitivity. Specifically, a $\sim$0.1 keV$_{nr}$ threshold is
expected from a liquid argon (LAr) bubble chamber. The scintillation
signal also helps with improving the timing of the detector, otherwise
limited to $\sim25 \mu$s through the acoustic emission from rapidly
expanding bubbles \cite{eric,cirte}.

Besides the mentioned ultra-low energy threshold and insensitivity to
ERs, bubble chambers exhibit a few additional features of interest for
CE$\nu$NS detection at the ESS: {\it i}) a 3-D event reconstruction of
order 1 mm permits to distinguish neutrino interactions from neutron
backgrounds, due to their different spatial distribution. {\it ii})
Suitable (but non-scintillating) targets containing light elements
such as C, F, and Br exist, and have been tested within the COUPP,
PICASSO, and PICO dark matter searches. For these the CE$\nu$NS cross
section is small (e.g., C$_{3}$F$_{8}$, Fig.\ \ref{fig:rates}),
increasing the difficulty of a measurement. The extreme insensitivity
to ERs of order 10$^{-9}$ \cite{baxter} seems crucially important in
order to achieve an eventual CE$\nu$NS detection in light
targets. {\it iii}) A dominant $^{39}$Ar ER background limits the
performance of conventional LAr detectors during CE$\nu$NS
measurements \cite{rex,rex2}. Use of LAr within a scintillating bubble
chamber bypasses this issue, due to the intrinsic insensitivity to
beta emitters. {\it iv}) The ultra-low 0.1 keV$_{nr}$ NR threshold
expected from a scintillating LAr bubble chamber makes it highly
sensitive to deviations in expected CE$\nu$NS signal rate induced by a
finite neutrino magnetic moment (Sec.~\ref{sec:pheno}).

\begin{figure}[!htbp]
\centering  
\includegraphics[width=1\columnwidth]{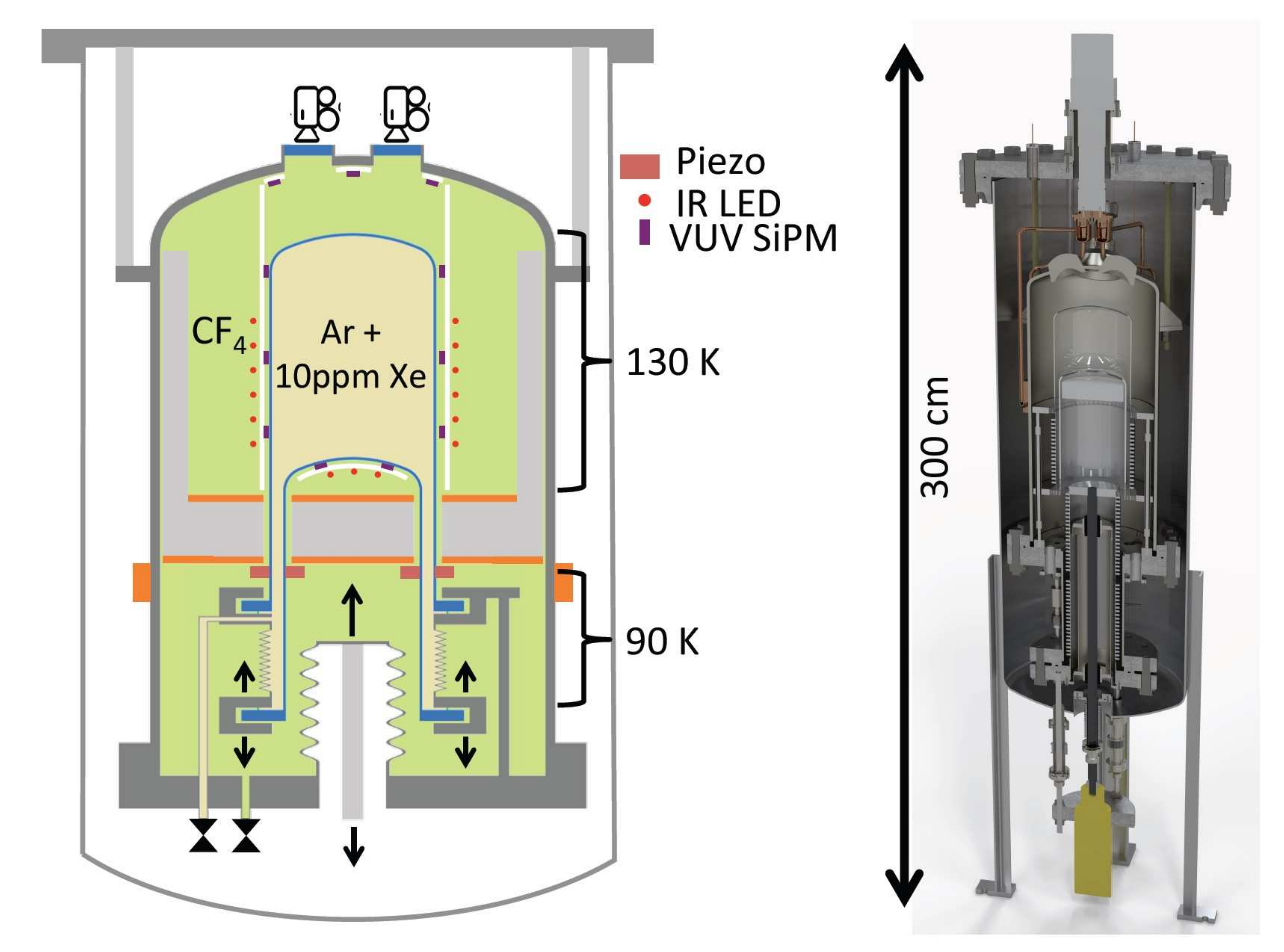}
\caption{\label{fig:bc} Schematic and model of a 10-kg LAr bubble
  chamber under construction, showing pressure and temperature
  control, bubble imaging, and scintillation detection scheme. The
  design is of a ``right side up" type, as in \cite{eric} and the
  upcoming PICO-40 and PICO-500 chambers at SNOLAB. Rendering by
  Fermilab's PPD/Mechanical Engineering Department.}
\end{figure}

PICO experimentation has demonstrated that the
thermodynamically-defined threshold for bubble formation is
well-defined and predictable \cite{pico,cirte}. This allows to obtain
spectral information by scanning operating parameters (typically
pressure), at the expense of an increased exposure. The NR background
adopted in Table~\ref{tab:detectors} is derived from unshielded runs
of a 2 kg CF$_{3}$I bubble chamber \cite{coupp} at 6 m.w.e., including
the simulated effect of 40 cm of neutron moderator shielding and 20 cm
of recompression/refrigerant fluid on environmental neutrons at this
overburden.

Presently, two 10-kg scintillating bubble chambers are under
construction at Northwestern University and Fermilab
(Fig.\ \ref{fig:bc}), one dedicated to dark matter searches, the
second intended for CE$\nu$NS experimentation. The time foreseen until
their full commissioning is a good match to the ESS start-up schedule.

\section{Physics Reach}
\label{sec:pheno}
As mentioned in Sec.~\ref{sec:intro}, a precision measurement of
CE$\nu$NS provides a direct probe to both SM and beyond the standard
model (BSM) physics.  Paradigmatic examples of the former are the
determination of the weak mixing angle at very low momentum transfer
\cite{wma1,wma2,wma3}, and the study of nuclear
structure~\cite{nst1,nst2,nst3,nst4}.  The program of BSM exploration
with CE$\nu$NS is broad (see
\cite{nsi1,nsi2,nsi3,nsi4,nsi5,nsi6,nsi7,nsi8,nsi9,nsi10,em1,em2,em3,em4,ste1,dm1,dm2,dm3,Khan:2019cvi,Cadeddu:2019eta,
  Papoulias:2019txv, Giunti:2019xpr} for an incomplete list) being
most sensitive to a variety of scenarios leading to modified neutrino
interactions with nuclei -- in particular at low momentum transfer --
but extending also to the production of new light neutral states, and
sterile neutrino searches, among others.

On this front, although part of the rationale for using a variety of
targets at the ESS is to demonstrate the N$^{2}$ dependence of the
CE$\nu$NS cross-section \cite{scholberg}, much more interesting is the
synergy in constraining the BSM physics that the use of multiple
targets can bring, and the advantages that some of these technologies
provide for specific aspects of the phenomenology explored. An
additional argument is that of redundancy: any observed deviations
from SM CE$\nu$NS predictions will require independent tests. In this
respect of anomaly confirmation, xenon and CsI targets provide a
rather unique combination of nearly-identical response to CE$\nu$NS
(Fig.\ \ref{fig:rates}), while relying on fundamentally different
techniques, subject to systematics not in common.

In what follows we illustrate the potential sensitivity of a CE$\nu$NS
experiment at the ESS to SM and BSM physics, for a variety of target
nucleus and detection technologies, with three characteristic
examples: NSI, the weak mixing angle and neutrino charge radius, and
an anomalous neutrino magnetic moment.

\subsection{Parameters and assumptions used in calculations}

For neutrino energies in the 50~MeV range, as it is the case for
neutrinos produced from pion DAR, the coherence condition is satisfied
for medium-sized nuclei. In the SM the differential cross section for
CE$\nu$NS on a nucleus with $Z$ protons and $N$ neutrons
reads~\cite{freedman}:
\begin{equation}
\label{eq:xsec-SM}
\frac{d\sigma_{SM} (T, E_\nu)}{dT } = \frac{G_F^2}{ 2 \pi}
\frac{\mathcal{Q}^2 (Z, N) }{ 4} F^2(Q^2) M \left(2 - \frac{M T }{
  E_\nu^2} \right),
\end{equation}
where $T$ is the recoil energy of the nucleus, $M$ is its mass,
$E_\nu$ is the incident neutrino energy, $G_F$ is the Fermi constant,
and $F$ is the form factor of the nucleus evaluated at the squared
momentum transfer of the process, $Q^2 = 2 M T$.  Here $\mathcal{Q}^2
\equiv 4 \left(Z g_{V,p} + N g_{V,n}\right)^2$, with $g_{V,p} = 1/2 -
\sin^2\theta_w$ and $g_{V,n} = -1/2$ being the weak charges of the
proton and the neutron, respectively. In our calculations, the weak
mixing angle has been set to its value at zero momentum transfer
$\sin^2\theta_w = 0.23867$, following
Refs.~\cite{pdg,Erler:2017knj}. As for the form factors, most of them
are readily available from Ref.~\cite{Horowitz:2003cz}. For molecules,
however, since these are not available we take the weighted average
between the form factors for the different nuclei. In the case of
C$_3$F$_8$, in the absence of a form factor for F we take the
corresponding one for O instead. In the case of CsI we use the Helm
form factor parametrization~\cite{Helm:1956zz}, using $s=0.9$~fm and
$R_n=4.83$~fm.

All sensitivity calculations in this work are obtained in nuclear
recoil energy space, after accounting for the effect of the QF on the
detector threshold and backgrounds.  The differential number of events
for the signal reads
\begin{equation}
\label{eq:dNdE}
\frac{dN}{dT} = \mathcal{N} \sum_{\alpha \equiv \nu_e, \nu_\mu,
  \bar\nu_\mu} \int dE_\nu \frac{d\sigma (T, E_\nu)}{dT}
\frac{d\phi_\alpha (E_\nu)}{dE_\nu} \, ,
\end{equation}
where $\mathcal{N}$ is a normalization constant that depends on the
number of protons on target, the neutrino yield per proton, the
detection efficiency (or acceptance), the mass of the detector, and
its distance to the source.  Unless otherwise stated, a common set of
assumptions apply to this normalization constant, for all detector
configurations considered in this work:
\begin{enumerate}
\item the detector distance to the ESS target is set to 20~m;
\item the detector signal acceptance is assumed to be a step function
  at threshold, with a conservative 80\% acceptance;
\item the running time is restricted to a total of 3 years. We think
  that this is reasonable, given the much longer running times
  envisioned for the physics program at the
  ESS~\cite{diff}. Increasing it beyond this value may lead to an
  improvement in our results only if the systematic errors can be
  reduced with respect to our assumed benchmark values outlined below.
\end{enumerate}

In what respects the expected backgrounds, they can be divided into
three classes: (\textit{i}) steady-state backgrounds (dominated by
cosmic ray interactions or by their by-products inside or in the
surroundings of a radioclean detector); (\textit{ii}) beam-related
backgrounds, produced by neutrons escaping the target and reaching the
detector; and (\textit{iii}) neutrino-induced neutrons. While the
latter is irreducible, it has been shown that its contribution to the
total event rate at the SNS is very small \cite{science,bjorn}, and
therefore will be neglected here. Beam-related backgrounds have also
been neglected (awaiting confirmation from ESS neutron background
studies), assuming that similar levels to those in \cite{science} or
shown in Fig.\ \ref{fig:ppc} are achievable. Therefore, our main
backgrounds are expected to come from the steady-state
contribution. For simplicity, this is assumed to follow a uniform
distribution in recoil energy.

%%%%%%%%%%%%%%%%%%%%%%%%%%%%%%
%\begin{widetext}
\begin{table*}[!htbp]
  \renewcommand{\arraystretch}{1.4} \centering
{\scriptsize
 \begin{tabular}{|@{\hspace*{2pt}}c@{\hspace*{2pt}}|@{\hspace*{2pt}}c@{\hspace*{2pt}}|@{\hspace*{2pt}}c@{\hspace*{2pt}}|@{\hspace*{2pt}}c@{\hspace*{2pt}}|@{\hspace*{2pt}}c@{\hspace*{0pt}}|@{\hspace*{2pt}}c@{\hspace*{2pt}}|@{\hspace*{2pt}}c@{\hspace*{2pt}}|@{\hspace*{2pt}}c@{\hspace*{2pt}}|@{\hspace*{2pt}}c@{\hspace*{2pt}}|@{\hspace*{2pt}}c@{\hspace*{2pt}}|}
 		\hline Detector Technology & Target & Mass &
                Steady-state & E$_{th}$ & QF & E$_{th}$ & $\frac{\Delta E}{E}$
                (\%) & E$_{\mathrm{max}}$ & CE$\nu$NS $\frac{\rm NR}{\rm yr}$ \\ &
                nucleus & (kg) & background & (keV$_{ee}$) & (\%) &
                (keV$_{nr}$) & at E$_{th}$ & (keV$_{nr}$) & @20m,
                $>$E$_{th}$ \\ \hline
                Cryogenic scintillator & CsI &
                22.5 & 10 ckkd& 0.1 & $\sim$10 \cite{csiqf} & 1 & 30 &
                46.1 & 8,405 \\
                Charge-coupled device & Si & 1 & 1
                ckkd & 0.007 & 4-30 \cite{alvaro} &
                0.16 & 60 & { 212.9} & 80 \\
                High-pressure gaseous TPC &
                Xe & 20 & 10 ckkd & 0.18 & 20 \cite{gas_xenon_nuclear}
                & 0.9 & 40 & 45.6 & 7,770 \\
                p-type point contact HPGe
                & Ge & 7 & 15 ckkd & 0.12 & 20 \cite{geqf} & 0.6 & 15
                & 78.9 & 1,610 \\
                Scintillating bubble chamber & Ar &
                10 & 0.1 c/kg-day & - & - & 0.1 & $\sim$40 & 150.0 &
                1,380 \\
                Standard bubble chamber & C$_{3}$F$_{8}$ & 10
                & 0.1 c/kg-day & - & - & 2 & 40 & 329.6 & 515
                \\ \hline
  \end{tabular}
}  
	\caption{\label{tab:detectors} Summary of detector properties,
          maximum recoil energies considered and expected signal and
          background rates used in our sensitivity calculations.
          Backgrounds listed do not include the
          $4\times10^{-2}$ reduction by the ESS duty factor. The
          germanium QF in \cite{geqf} will be revisited in an upcoming
          publication. A rapid dependence of the silicon QF on NR
          energy \cite{alvaro,javier2} is adopted. The concept of QF is
          ill-defined for bubble chambers (all NR energy is available
          for nucleation). Their background is integrated above
          nucleation threshold (in counts per kg and day).
          Backgrounds for semiconductors are in per keV$_{ee}$, while
          for cryogenic CsI and pressurized xenon this is given in per
          keV$_{nr}$, and in both cases are given in counts per keV,
          kg and day (ckkd). We conservatively adopt the background at E$_{th}$, which is typically maximal, for all higher energies.}

\end{table*}
%\end{widetext} 
%%%%%%%%%%%%%%%%%%%%%%%%%%%%%%
%
Table~\ref{tab:detectors} summarizes the detector properties,
steady-state background, and QF values assumed for all detectors
considered in this work. With the exception of bubble chambers (see
below), a Gaussian energy smearing is applied, with a width that
depends on the recoil energy as: $\sigma(T) = \sigma_0
\sqrt{T/T_{thres}}$, where $\sigma_0$ is the energy resolution at the
detection threshold ($T_{thres}$), which can be readily extracted from
Table~\ref{tab:detectors}.  For each detector, the recoil energy bin sizes are
chosen so that the width of each bin is twice the energy
resolution at its center.
We consider all the kinematically available range (determined by the
condition $T \lesssim 2 E_\nu^2 / M $), for all detector configurations.

While most detectors under consideration have excellent energy
reconstruction capabilities, for bubble chambers only the total event
rates are used in the calculations: no bins or energy smearing are
used to compute the event rates. For these detectors, however, it is
possible to predict and adjust the detection threshold with high
precision \cite{pico,cirte}.  Splitting the running time of a bubble chamber into two periods (with different detection thresholds) would lead to an
increased sensitivity to NP scenarios which manifest at low
recoils but remain unnoticed for events in the higher energy part of
the spectrum, as we will see in Sec.~\ref{sec:mu}.

In order to determine the sensitivity to a NP model
characterized by a set of parameters $\left\{ \varepsilon
\right\} $, a binned $\chi^2$ is built. 
Systematic uncertainties are implemented using the pull method, and
assumed to be fully correlated among different energy bins while totally
independent for signal and background. Thus we introduce two nuisance
parameters $\xi_{sig}$ and $\xi_{bg}$ to parametrize these independent
normalization uncertainties. Altogether
\begin{eqnarray}
\chi^2\big( \{\varepsilon\}\big) = min_{\xi}
&\Big[&
\chi^2\big(\{\varepsilon\},\xi\big) + 
\left(\frac{\xi_{sig}-1}{\sigma_{sig}}\right)^2 + \nonumber \\ &&
\left(\frac{\xi_{bg}-1}{\sigma_{bg}}\right)^2 \Big] \,  ,
\label{eq:chi2}
\end{eqnarray} 
where
\begin{equation}
\chi^2 (\left\{ \varepsilon \right\},\xi ) = \sum_i 2 \Big[ N_i (\{
  \varepsilon \},\xi ) - \bar N_i
 + \bar N_i \ln \left(\frac{\bar
    N_i }{N_i (\{ \varepsilon\},\xi ) }\right) \Big] .
\end{equation}
Here $N_i (\{ \varepsilon \},\xi )$ stands for the event rate/s in
the $i$-th energy bin
(adding up signal and background rates each with its corresponding
normalization factor) predicted by the model that is
being tested, while $\bar N_i$ stands for the event rates expected
in that bin from the combination of signal and background in the SM.
In Eq.~\eqref{eq:chi2}, $\sigma_{sig}$ and $\sigma_{bg}$ are  signal and background normalization
uncertainties. In the results presented here  a 10\%
systematic uncertainty has been assumed for the signal prediction. For
reference, the corresponding value at the COHERENT experiment
currently exceeds 20\%, after adding all corresponding
uncertainties in quadrature. However, the largest contributor to such
error is the systematic error on the QF, which according to the
results from Ref.~\cite{csiqf} can be reduced down from a 18\% to
approximately a 6\%. Therefore, we deem a 10\% representative of the
cumulative uncertainties in QF, neutrino flux, nuclear form factor,
and signal acceptance by the time data taking starts. While the
steady-state background is in principle subject to sizable systematic
uncertainties, it will be efficiently measured using beam-off data, as
demonstrated at COHERENT \cite{science,bjorn}. Therefore, this systematic uncertainty is
set to 1\% in our calculations for all detectors under consideration.
{ It is important to stress that, given the large statistics
predicted in all considered detectors, most of the sensitivity results presented
below are limited by systematics within our simplified treatment and 
conservative assumptions about their uncertainties.
They are, therefore, subject to improvements by dedicated campaigns allowing
a better characterization of the systematic uncertanties in each
detector.}
  
\subsection{Non-Standard neutrino Interactions}
\label{sec:nsi}

From a completely model-independent approach, a useful parametrization
of the possible effects of NP at low energies is through the
addition of higher-dimensional operators to the SM Lagrangian, that
respect the SM gauge group. At $d=5$, the only operator that can be
built using just SM fields is the Weinberg operator~\cite{Weinberg:1979sa}, which
coincidentally gives rise to neutrino masses. At $d=6$, the allowed
set of operators includes four-fermion operators affecting neutrino
production, propagation, and detection processes.  For example,
operators of the form
\begin{equation}
\label{eq:nsi-cc}
2\sqrt{2} G_F \epsilon_{\alpha\beta}^{ff',P} 
(\bar\nu_\alpha \gamma_\mu P_L \ell_\beta)(\bar f' \gamma^\mu P f ) 
\end{equation}
would induce non-standard charged-current (CC) production
and detection processes for neutrinos of flavor $\alpha$,
while operators such as
\begin{equation}
\label{eq:nsi-nc}
2\sqrt{2} G_F \epsilon_{\alpha \beta}^{f,P} 
(\bar\nu_\alpha \gamma_\mu P_L \nu_\beta)(\bar f \gamma^\mu P f)
\end{equation}
would lead to flavor-changing neutral-current (NC) interactions of
neutrinos with other fermions (if $\alpha \neq \beta$), or to a
modified NC interaction rate with respect to the SM expectation (if
$\alpha = \beta$). In Eqs.~\eqref{eq:nsi-cc} and~\eqref{eq:nsi-nc},
$f$ and $f'$ refer to SM fermions, $\ell$ refers to a SM charged lepton
and $P$ can be either a left- or a right-handed projection operator
($P_L$ or $P_R$, respectively).

While CC NSI are severely constrained by the study of CC processes,
such as meson and muon decays, constraining NC NSI is a much more
challenging task.  This is so because of the uncertainties involved in
computing neutrino-nucleus interactions, and the experimental
difficulties in measuring NC cross sections precisely. In fact, the
best constraints available in the literature for these operators come
from global fits to oscillation data, which are very sensitive to
modifications in the effective matter potential felt by neutrinos as
they propagate in a medium \cite{nsi7}. Consequently they can bound
vector NSI
($\epsilon_{\alpha\beta}^{f,V}\equiv\epsilon_{\alpha\beta}^{f,L}+\epsilon_{\alpha\beta}^{f,R}$)
and, since they are due to a totally coherent effect, these bounds
extend to NSI induced even by ultra light mediators ($M_{\rm
  med}\gtrsim 1/R_{\rm Earth}\sim {\cal O}(10^{-12})\,{\rm (eV)}$).
However, while oscillation experiments are sensitive to all
flavor-changing NSI, they are only sensitive to differences between
the diagonal NSI parameters in flavor space \cite{nsi7,nsi10}. This
leads to the appearance of new degeneracies involving standard
oscillation parameters and NSI operators, such as the so-called
generalized mass ordering degeneracy~\cite{Coloma:2016gei,
  Gonzalez-Garcia:2013usa, Bakhti:2014pva}, with important
consequences for the upcoming generation of long-baseline neutrino
oscillation experiments.

Conversely CE$\nu$NS
experiments at spallation sources
allow to constrain two
of the three flavor-diagonal coefficients, since the neutrino flux
contains both muon and electron neutrinos.
In brief, the
differential number of signal events in the presence of NC NSI can be
obtained from Eqs.~\eqref{eq:dNdE} and~\eqref{eq:xsec-SM} with
the effective coupling $\mathcal{Q}^2$ in Eq.~\eqref{eq:xsec-SM} modified as~\cite{Barranco:2005yy}:
\begin{eqnarray}
\label{eq:Qalpha-nsi}
{\mathcal{Q}_{\alpha}^2}^{(\epsilon)}& = & 4 \left[Z \left(g_{V,p} +
  2\epsilon_{\alpha\alpha}^{u,V} + \epsilon_{\alpha \alpha}^{d,V}
  \right) \right.+ \nonumber \\ 
  & & + \left. 
  N \left( g_{V,n} + \epsilon_{\alpha\alpha}^{u,V} +
  2\epsilon_{\alpha \alpha}^{d,V} \right) \right]^2  \\ & + &
4 \sum_{\beta \neq \alpha} \left[ Z ( 2 \epsilon_{\alpha \beta}^{u,V}
  + 2 \epsilon_{\alpha \beta}^{d,V}) + N (\epsilon_{\alpha\beta}^{u,V}
  + 2 \epsilon_{\alpha\beta}^{d,V}) \right]^2 \, , \nonumber
\end{eqnarray}
where we explicitly note that its value will now generally depend on
the incident neutrino flavor $\alpha$. 
Notice that both oscillation and CE$\nu$NS experiments are
sensitive to the vector  NSI.

Thus, the complementarity between neutrino oscillation and coherent
scattering data is evident: the addition of coherent scattering data to
the global fits from oscillation breaks the degeneracies involving
flavor-diagonal NSI, thanks to the direct measurement of the
neutrino-nucleus interactions for both electron and muon
neutrinos~\cite{Coloma:2016gei,nsi2}. Furthermore, while in principle
incoherent neutrino scattering at higher energies is also sensitive to
these operators, when the NSI is induced by light mediators
the sensitivity of these scattering experiments (such as
NuTeV~\cite{Zeller:2001hh} or CHARM~\cite{Dorenbosch:1986tb}) becomes
limited by the larger minimum momentum transfer required in the
inelastic scattering detection~\cite{Farzan:2015doa, Farzan:2015hkd, Coloma:2016gei, nsi10}. 

With all these considerations, in what follows, we focus on the
determination of the flavor-diagonal NSI coefficients,
$\epsilon_{\alpha\alpha}^{f,V}$ ($f=u,d$), although it should be kept
in mind that coherent neutrino scattering is also sensitive to all the
off-diagonal NSI operators as well, and competitive bounds should also
be expected for those. 

%%%%%%%%%%%%%%%%%%%%%%%%
\begin{figure*}[ht!]
\centering  
\includegraphics[width=0.85\textwidth]{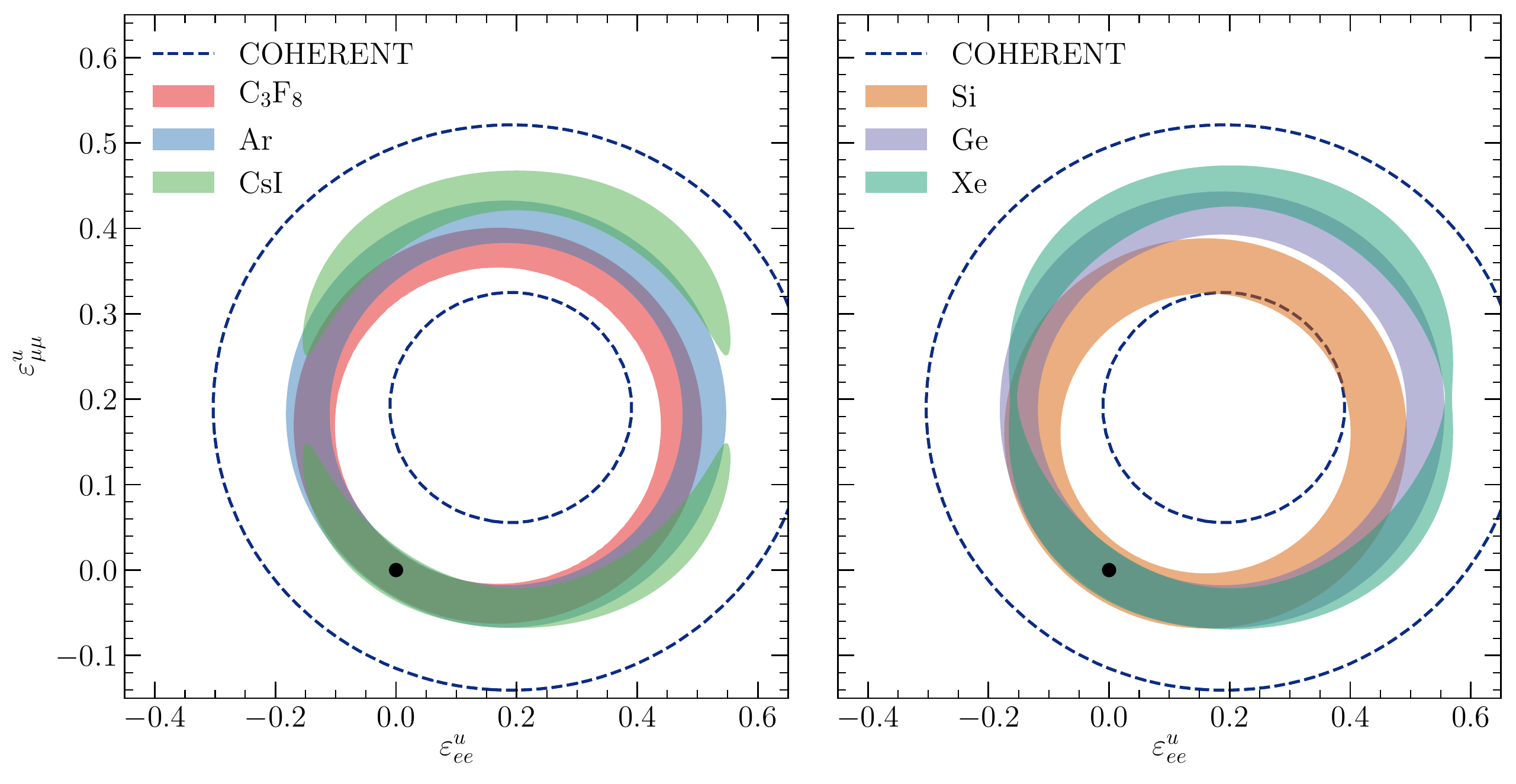}
\caption{\label{fig:nsi-alldets} Expected allowed regions in the
  $(\epsilon_{ee}^{u,V} ,\epsilon_{\mu\mu}^{u,V})$ plane at the 90\%
  confidence level (C.L.) for two degrees of freedom (d.o.f.). The
  different regions correspond to the expected results for the
  different detectors listed in Table~\ref{tab:detectors}, as
  indicated by the legend. In all cases, the simulated data has been
  generated for the SM (that is, setting all the operator coefficients
  to zero), and the results are then fitted assuming NSI. For
  simplicity, the rest of the NSI parameters not shown in the figure
  have been assumed to be zero. For comparison, the dashed lines show
  the allowed regions at 90\% CL in this plane, as obtained in Ref.~\cite{nsi2} from an analysis of current data from the COHERENT
  experiment~\cite{science}, see text for details. }
\end{figure*}
%%%%%%%%%%%%%%%%%%%%%%%%
Figure~\ref{fig:nsi-alldets} shows our results on the expected allowed
regions at 90\% CL in the plane $(\epsilon_{ee}^{u,V} ,
\epsilon_{\mu\mu}^{u,V})$ for the six detectors under consideration.
In this figure for simplicity, we have assumed that the NSI take place
only with up-type quarks; however, similar results are obtained if the
NSI are assumed to take place with down-type quarks instead. For
illustration we show as well the 90\% CL allowed region from the
analysis of the total event rate observed at the COHERENT experiment
in Ref.~\cite{nsi7}, following the prescription provided in
Ref.~\cite{science} to perform a fit to NSI using the total event
rates. In principle, adding the timing and energy information provided
in Ref.~\cite{Akimov:2018vzs} can help to tighten their constraints to
some degree~\cite{Cadeddu:2019eta,Giunti:2019xpr}; however, the final
result is subject to uncertainties in the treatment of the background
and systematic errors assumed.

The different areas in the two panels in Fig.~\ref{fig:nsi-alldets}
correspond to the results obtained with the detector configurations
listed in Table~\ref{tab:detectors}. As seen in the figure, in most
cases the shape of the allowed regions is an ellipse in this plane.
This can be easily understood as follows.  From
Eqs.~\eqref{eq:dNdE},~\eqref{eq:xsec-SM}, and ~\eqref{eq:Qalpha-nsi}
one can trivially compute the expected total number of events (adding
up the contributions from the three components of the beam) in each
bin as a function of the two NSI coefficients in each.  Requiring that
the NSI-induced correction is of the same relative size in all bins
and that the total number of events is compatible with the SM
expectation, it is straightforward to show that the allowed confidence
regions in the plane $\epsilon_{ee}^u - \epsilon_{\mu\mu}^u $ obey the
equation of an
ellipse:
\begin{equation}
\label{eq:ellipse}
\left[ R + \epsilon_{ee}^{u,V} \right]^2 + 
2 \left[ R + \epsilon_{\mu\mu}^{u,V}\right]^2 = 3 R^2
\end{equation}
where $R \equiv \frac{Z g_{V,p} + N g_{V,n}}{2Z + N} $ only depends on
the target nucleus and the SM weak couplings to protons and
neutrons. In the SM, given that $g_{V,p} \ll g_{V,n}$ this constant can be
safely approximated to $ R \simeq g_{V,n} / (2 r + 1)$, where $r\equiv
Z/N$ is the ratio of protons to neutrons in the nucleus. 
From Eq.~\eqref{eq:ellipse} it follows that the shape of the allowed
confidence regions in this plane will be very similar for different
target nuclei as long as they have a similar value of $r$.
For reference
Table~\ref{tab:nuclei} summarizes the values of $Z$, $N$, $r$, and the
nuclear masses assumed for different nuclei.
%%%%%%%%%%%%%%%%%%%
\begin{table}
\centering  
\begin{tabular}{| c | cccc |}
\hline
Nucleus & $Z$ & $N$ & $r$ & $M$(a.m.u.) \\ \hline  
 $^{132}$Xe & 54 &  78 & 0.69  & 131.29   \\ 
 $^{40}$Ar & 18 & 22 & 0.81  & 39.95 \\
 $^{72}$Ge & 32 &  40 & 0.8 & 75.92 \\
 $^{28}$Si & 14 &  14 & 1.0 & 27.98 \\
 $^{12}$ C & 6 & 6 & 1.0 & 12.01  \\
 $^{19}$F & 9 & 10 & 0.9 & 19.00 \\
 $^{133}$Cs &  55 & 78 & 0.71 & 132.91 \\
 $^{127}$I & 53 & 74 & 0.72 & 126.90 \\ 
 $^{20}$Ne & 10 &  10 & 1.0 & 20.18 \\
 \hline
\end{tabular}
\caption{Main properties of the nuclei for the different target nuclei
  considered in this work. The different columns indicate the isotope
  considered, together with the number of protons and neutrons, the
  ratio between them $r$, and the value of the nuclear mass in atomic
  mass units (a.m.u.). For the detectors using CsI and C$_3$F$_8$ we
  take the weighted average between the two elements in the molecule.
\label{tab:nuclei}}
\end{table}
%%%%%%%%%%%%%%%%%%%

As seen in Fig.~\ref{fig:nsi-alldets}, the allowed regions are in good
agreement with Eq.~\eqref{eq:ellipse}, for most of the detectors under
consideration. However, from the figure we also see that for some
detectors, in particular for the CsI target (and also in part for Xe
target), the degeneracy in the allowed region in the
$(\epsilon_{ee}^{u,V} , \epsilon_{\mu\mu}^{u,V})$ plane implied by
Eq.~\eqref{eq:ellipse} is partly broken.  This is so because
Eq.~\eqref{eq:ellipse} has been obtained under the approximation of a
constant -- flavour- and energy-independent -- shift
of the event rates in all bins. Clearly the degeneracy will be
broken if somehow the experiment is capable of discriminating between muon
and electron neutrino flavors at some level.
A possibility to do this, is through
the addition of timing information, which allows to distinguish between
the prompt component of the beam (which contains just $\nu_\mu$) and
the delayed component (which contains a mixture of $\nu_e$ and
$\bar\nu_\mu$). Unfortunately, due to the very long proton pulses this would not be possible at the ESS source.

\begin{figure}[ht!]
\centering  
\includegraphics[width=1\columnwidth]{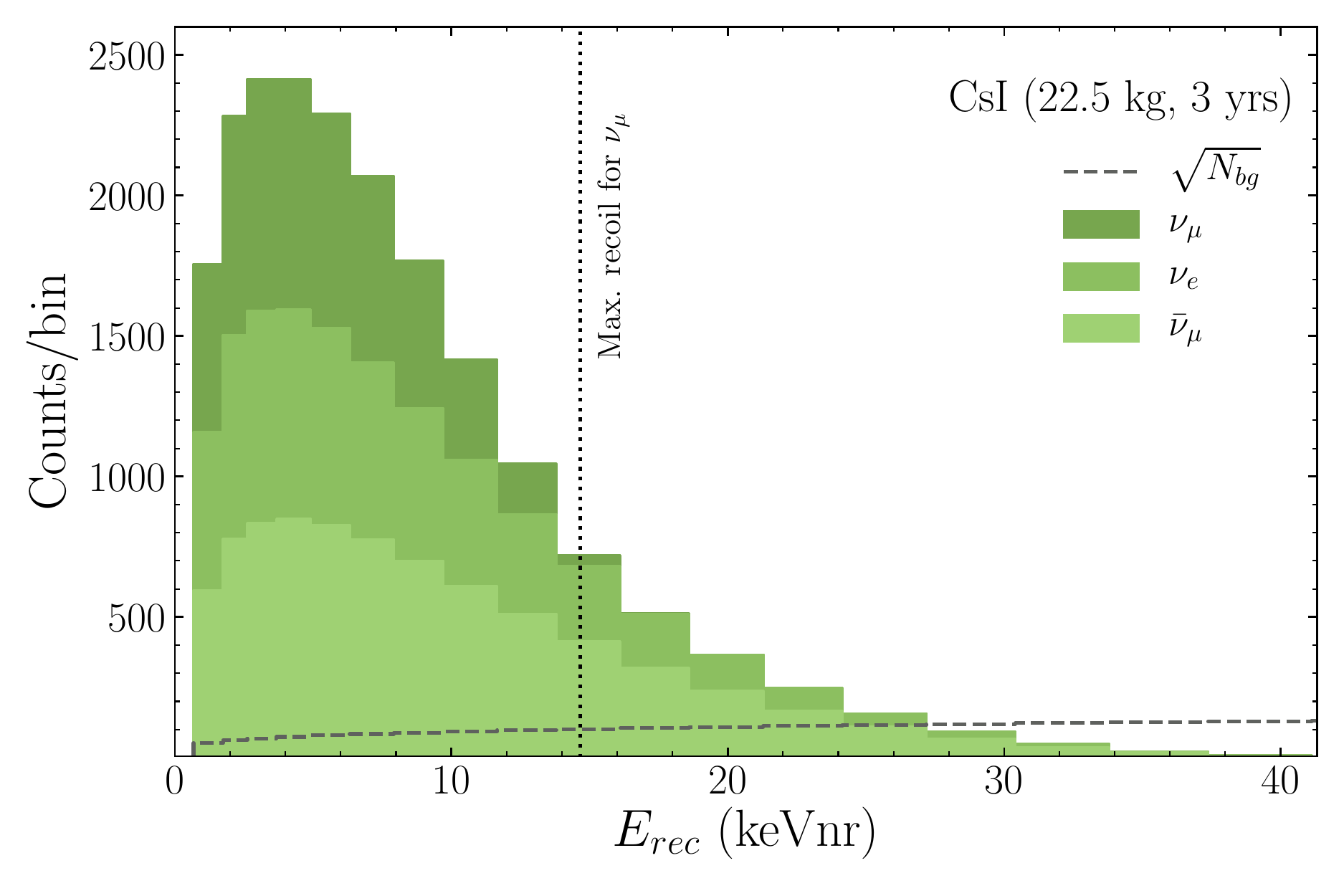}
\caption{\label{fig:events-CsI} Expected event rates per bin in
  nuclear recoil energy, for the CsI detector. The contributions from
  the scattering of the different beam components are shown separately
  by the shaded histograms, as indicated by the legend. For
  comparison, the square root of the number of background events in
  each bin is also shown by the dashed histogram lines. The vertical
  dotted line indicates the maximum recoil energy allowed by a
  neutrino with energy $E_\nu = 29.8$~MeV, that is, the energy of the
  monochromatic prompt $\nu_\mu$ flux. }
\end{figure}

One must notice, however, that the prompt signal is also characterized by a
lower neutrino energy ($E_{\nu_\mu} \sim 30$~MeV). Therefore, it should be
possible to distinguish its contribution using a detector with good
energy resolution that allows to observe not only the bulk of events
at low energies (which receives equal contributions from the three
components of the beam) but also the tail at high recoil energies,
above the maximum recoil allowed for the prompt signal. For large
enough statistics, this would allow to obtain partial flavor
discrimination, by comparing the event rates below and the maximum
recoil allowed for the prompt flux. For illustration, we show in
Fig.~\ref{fig:events-CsI} the expected event rates, where the
contribution per flavor is shown separately. As shown in this figure,
above the maximum recoil energy allowed for the prompt component the
event rates are given almost exclusively by $\nu_e$ and $\bar\nu_\mu$
scattering (albeit with a small contribution from $\nu_\mu$ in the
first bin, due to smearing by the energy resolution).
%%%%%%%%%%%%%%%%%
%%%%%%%%%%%%%%%%
Consequently in the case of detectors with high statistics, good energy resolution and no saturation, the ellipse is broken in this plane. This is the case for the CsI and Xe detectors, as
seen in Fig.~\ref{fig:nsi-alldets}.

\begin{figure}[ht!]
\centering  
\includegraphics[width=1\columnwidth]{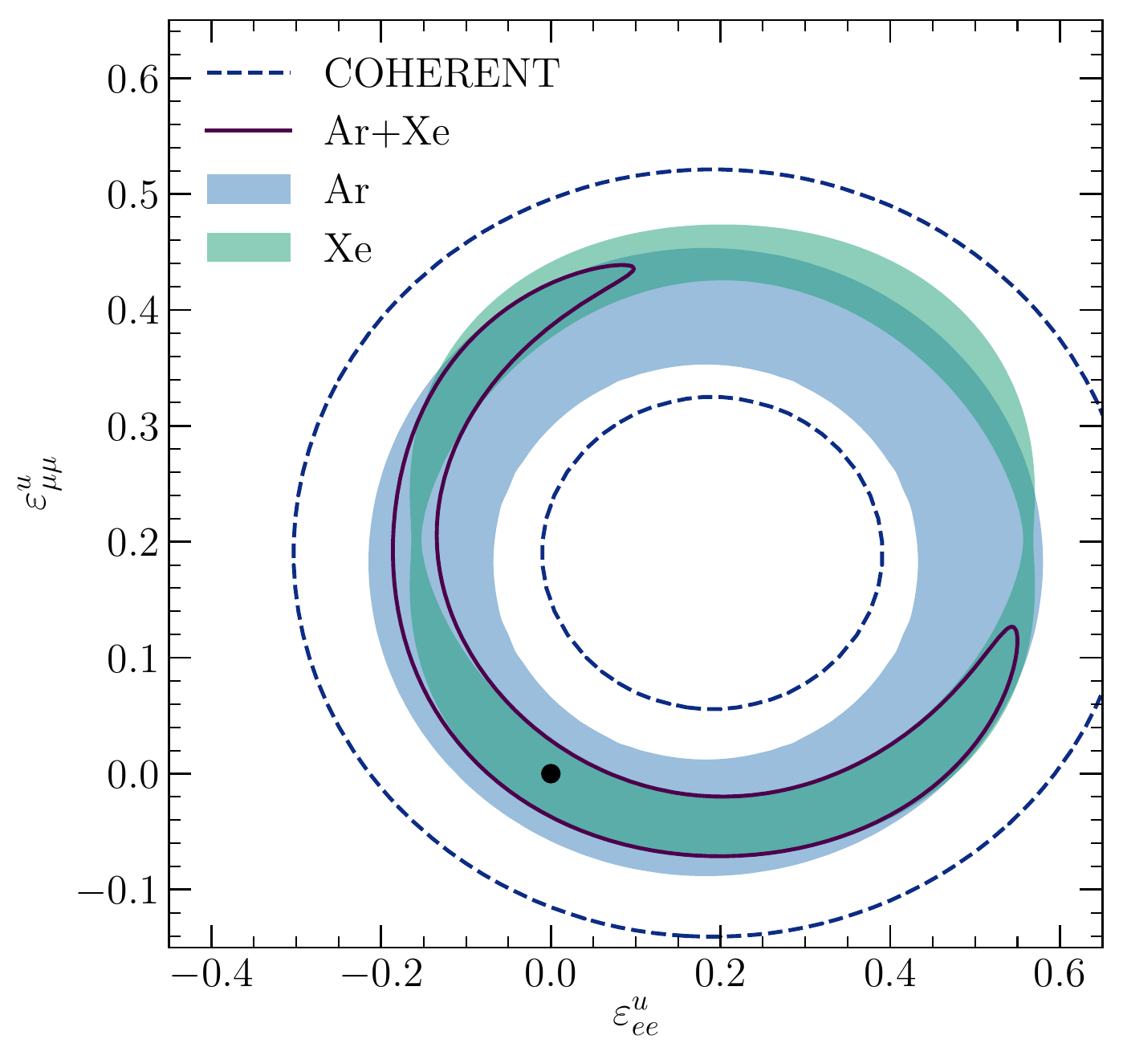}
\caption{\label{fig:next} Expected allowed regions in the
  $(\epsilon_{ee}^{u,V},\epsilon_{\mu\mu}^{u,V})$ plane at the
  90\% C.L. for 2 d.o.f, for the gas TPC detector operating with
  two different nuclei (separate runs, each of them for 3 years), as
  well as for a configuration where the detector is filled with each
  of the two gases during half of the total data taking period (1.5
  years running with $^{132}$Xe, 1.5 years with $^{40}$Ar). In all
  cases, the simulated data has been generated for the SM and the
  results are then fitted assuming NSI. For simplicity, the values for
  the rest of NSI parameters not shown in this figure have been set to
  zero.}
\end{figure}
%%%%%%%%%%%%%%%%%
Furthermore, from Eq.~\eqref{eq:ellipse} it is clear that an
alternative form of breaking this degeneracy is through the
combination of data obtained using different target nuclei, as long as
they have different values of $r$.  This is true even if only
information on the total event rates is available (without any
time nor energy information).  While the combination can be done using
different detectors among the possibilities listed in
Table~\ref{tab:detectors}, a more convenient option is available in the
case of the gas TPC, since the detector can operate not only with
xenon but with other noble gases as well (Ne, He, or Ar for instance).
As illustration of this possibility, we show Fig.~\ref{fig:next} the
expected sensitivity regions in this plane using the gas TPC
detector. In this case we show three different results: (\textit{i}) the
expected regions obtained using Xe as the target nucleus; (\textit{ii}) the
expected regions using Ar instead; and (\textit{iii}) the results obtained
using a combined run, where the detector uses Xe during the first half
of the data taking period and Ar during the other half. From the figure
we see how the combination of runs with the two selected nuclei
leads to a substantially improved sensitivity.

\subsection{Weak mixing angle and neutrino charge radii }
\label{sec:thetaW}

\begin{table*}
\renewcommand{\arraystretch}{1.4}
{\small
  \begin{tabular}{|@{\hspace*{+0.02cm}}c@{\hspace*{+0.02cm}}|@{\hspace*{+0.02cm}}c@{\hspace*{+0.02cm}}c@{\hspace*{+0.02cm}}c@{\hspace*{+0.02cm}}c@{\hspace*{+0.02cm}}c@{\hspace*{+0.02cm}}c@{\hspace*{+0.02cm}}c@{\hspace*{+0.02cm}}|@{\hspace*{+0.02cm}}c@{\hspace*{+0.02cm}}|}
\hline
& Ar & $\mathrm{C}_3\mathrm{F}_8$ & CsI & Ge & Si & Xe & Xe+Ar & COH-SNS 
\\  
\hline 
 $\sin^2 \theta_W$ 
& {\footnotesize
    $0.239^{+0.028}_{-0.022}$}
 & {\footnotesize$0.239^{+0.025}_{-0.020}$}
 & {\footnotesize$0.239^{+0.032}_{-0.026}$}
 & {\footnotesize$0.239^{+0.029}_{-0.024}$}
 & {\footnotesize $0.239^{+0.032}_{-0.029}$}
 & {\footnotesize$0.239^{+0.033}_{-0.026}$}
 & {\footnotesize$0.239^{+0.020}_{-0.029}$}
&$0.248\pm 0.094$\cite{Khan:2019cvi} 
\\
 \hline
 $\langle r^2_{ee}\rangle $ & [-65, 20] & [-58, 18] & [-67, 16] & [-67, 20] &
 { [-54, 18]} & [-70, 17] & [-55, 20] & [-65, 6] \cite{Cadeddu:2019eta}
\\
$\langle r^2_{\mu \mu}\rangle$ & [-51, 7] & [-46, 6] & -[59, 7] & [-54, 7] & [-43, 6.5] &
  [-60, 7.5] & [-28, 7] & [-60, 10] \cite{Cadeddu:2019eta}
\\
$|\langle r^2_{e \mu}\rangle|$ & $<$ 15 & $<$ 12 & $<$ 21 & $<$ 17 & $<$ 11 & $<$ 21 & $<$ 17  & $<$35 \cite{Cadeddu:2019eta}
\\ \hline 
$\mu_{\nu_\mu}$ &  $<$ 9 & $<$ 11 & $<$ 9 & $<$ 7 & $<$ 6 & $<$ 9 &
$<$ 10
& $<$31 \cite{Cadeddu:2019eta}
\\
\hline
\end{tabular}}
\caption{Allowed ranges at 90\% C.L. for the weak mixing
  angle (given as best fit $\pm 1.64 \sigma$), neutrino charge radii for three flavour
  projections  (in units of $10^{-32}$ cm$^2$, and after
  marginalizing over the other two flavour projections), and 
  the $\nu_\mu$ magnetic moment (90\% CL upper bound in units of
  $10^{-10} \mu_{\rm B}$).} 
\label{tab:macrotable}  
\end{table*}  

The weak mixing angle is a fundamental parameter in the SM. While
its value has been precisely measured at high
energies in collider experiments, its determination at low energy
is a challenging task from the experimental point of view. At low
energies, it can be determined from measurements of parity violation
in Cs atoms~\cite{Wood:1997zq,Dzuba:2012kx},
the parity-violating asymmetry in Moller
scattering~\cite{Anthony:2005pm}, deep inelastic scattering of polarized
electrons in deuteron \cite{Wang:2014bba},
and neutrino scattering on
nuclei~\cite{Zeller:2001hh}. While most of these measurements
seem to agree well within error bars, the NuTeV
result~\cite{Zeller:2001hh}
shows a tension at the $3\sigma$ CL.

As seen in Eq.~\eqref{eq:xsec-SM}, the  weak mixing angle enters the
neutrino-nucleus coherent scattering
cross section through the value of $g_{V,p}$. Therefore, its effect
on the number of events is going to be much more subleading than for
NSI, and its impact on the observable number of
events will be a change in the normalization of the event sample
which, in this case, will be flavor-universal. However, while the weak mixing angle affects the coupling to protons it does
not affect the coupling to neutrons and, therefore, an enhanced
sensitivity is expected also in this case by combining results
obtained for nuclei with different proton-to-neutron ratios $r$.

Our results on the expected sensitivity for this parameter are shown in
Table~\ref{tab:macrotable} for different detectors under
consideration.  For reference, the current bounds derived in
Ref.~\cite{Khan:2019cvi} from current COHERENT results, are also included.
As seen in the table, any of the experiments considered here can lead
to an improvement on the determination of the weak mixing angle in
CE$\nu$NS by a factor ${\cal O} (3)$. { Notice also that
the sensitivity of all dectors to this parameter is comparable
because it is mostly limited by the assumed 10\% normalization uncertainty.}

The determination of the weak mixing angle is tightly related to the
sensitivity to the effective neutrino charge radius, $\rnu$, defined
as~\cite{Giunti:2014ixa}
\begin{equation}
\label{eq:rnu}
\rnu = 6 \frac{dF_\nu(q^2)}{dq^2} \bigg |_q^2 = 0 \, ,
\end{equation}
where $F_\nu$ is the electromagnetic form factor of the neutrino. The
inclusion of this form factor affects the scattering of neutrinos with
other charged particles in the SM, and effectively induces a shift in
the value of the effective mixing angle~\cite{Degrassi:1989ip, Vogel:1989iv, Kouzakov:2017hbc}, 
\begin{equation}
\label{eq:thetaW-rnu}
\sin^22\theta_w \to \sin^22\theta_w \left( 1 + \frac{1}{3}m_W^2\rnu
\right) \, ,
\end{equation}
where $m_W$ is the mass of the $W$ boson.

The SM prediction for the neutrino charge radii gives a value that
depends on the neutrino flavor $\alpha \equiv e,\mu,\tau$, as~\cite{Bernabeu:2000hf,Bernabeu:2002nw,Bernabeu:2002pd}:
\begin{eqnarray}
\label{eq:rnu-SM}
\langle r_{\nu_\alpha}^2 \rangle = \frac{- G_F}{2\sqrt{2} \pi^2}
\left[ 3 - 2 \ln \left( \frac{m_{\ell_\alpha}^2}{m_W^2}\right) \right]
\, ,
\end{eqnarray}
where $m_{\ell_\alpha}$ is the mass of the charged lepton of the same
flavor $\alpha$. Therefore, determining the weak mixing angle
precisely allows to obtain an upper limit for the value of the charge
radius of each neutrino flavor state. Numerically, however, these lie
in the range $[ -0.83, -0.3 ] \times 10^{-32}$~cm$^2$ and are
therefore very challenging to observe. 

However, in BSM scenarios the
charge radii may receive additional contributions and,
therefore, a measurement of their value well above the SM expectation
would be a clear signal of NP. Moreover in BSM models with
neutrino masses, due to the mismatch between the mass and flavor
bases, transition charge radii $\langle r_{\alpha\beta}^2 \rangle$ may be 
generated for the flavor states. In the most general case where
both diagonal and transition charge radii are considered, the
effective weak coupling is modified as~\cite{Cadeddu:2019eta,Kouzakov:2017hbc}:
\begin{equation}
\label{eq:Qalpha-radii}
{\mathcal{Q}^2_{\alpha}}^{(r)} = 4 \left[Z \left(g_{V,p} - Q_{\alpha \alpha}
  \right) + N g_{V,n} \right]^2 + 4 Z^2 \sum_{\beta \neq \alpha} |
Q_{\alpha\beta} |^2
\end{equation}
where $Q_{\alpha\beta}$ is defined as
\begin{equation}
Q_{\alpha\beta} \equiv \frac{2}{3} m_W^2 \sin^2\theta_w \langle
r_{\alpha\beta}^2 \rangle \, .
\end{equation}
Note that, unlike for NSI, in this case the modification to the
neutrino cross section will be proportional to the number of protons
in the nucleus and therefore the expected effect on the number of
events will be different. Moreover, in presence of transition charge
radii additional modifications to the SM cross section, proportional
to $Z^2$, are also expected. Finally in this
scenario we also expect, as in the case of the weak mixing angle, an improvement in 
sensitivity through the
combination of different target nuclei with different values of
$r$. This will greatly help to cancel the effect of systematic
uncertainties and to increase the sensitivity to the values of
$Q_{\alpha\beta}$.

Our results for the determination of the neutrino charge radii are
summarized in Table~\ref{tab:macrotable}, for the different detectors
under consideration. For reference, the current bounds derived from
COHERENT data in Ref.~\cite{Cadeddu:2019eta}, are also included.
As seen in the table, for any of the detectors considered, an improvement
with respect to the present results from COHERENT is expected regarding 
the sensitivity to the charge radius for $\nu_\mu$ (although they cannot 
reach the SM expectation), as well as the flavour transition charge radii. 

\subsection{Neutrino magnetic moment}
\label{sec:mu}

In presence of a neutrino magnetic moment, the SM cross section
receives an extra contribution coming from the exchange of a photon
with the nucleus. The differential cross section reads~\cite{Vogel:1989iv}:
\begin{equation}
\label{eq:xsec-mu}
\frac{d\sigma}{dT} = \frac{d\sigma_{SM}}{dT} + \frac{\pi \alpha^2}{
  m_e^2} \left[ \frac{1}{T} - \frac{1}{E_\nu} + \frac{T}{4 E_\nu^2}
  \right] Z^2 F_{em}^2(Q^2) \left( \frac{\mu}{\mu_B} \right)^2 ,  
\end{equation}
where $m_e$ is the electron mass, $\mu_B$ is the Bohr's magneton, and
$\alpha$ is the electromagnetic fine constant.
It should be noted that, since the effect due to a non-zero magnetic
moment is only relevant for very low recoils, the
electromagnetic form factor $F_{em}(Q^2)$ can be safely approximated
to one.

%%%%%%%%%%%%%%%%
\begin{figure}
\centering  
\includegraphics[width=1\columnwidth]{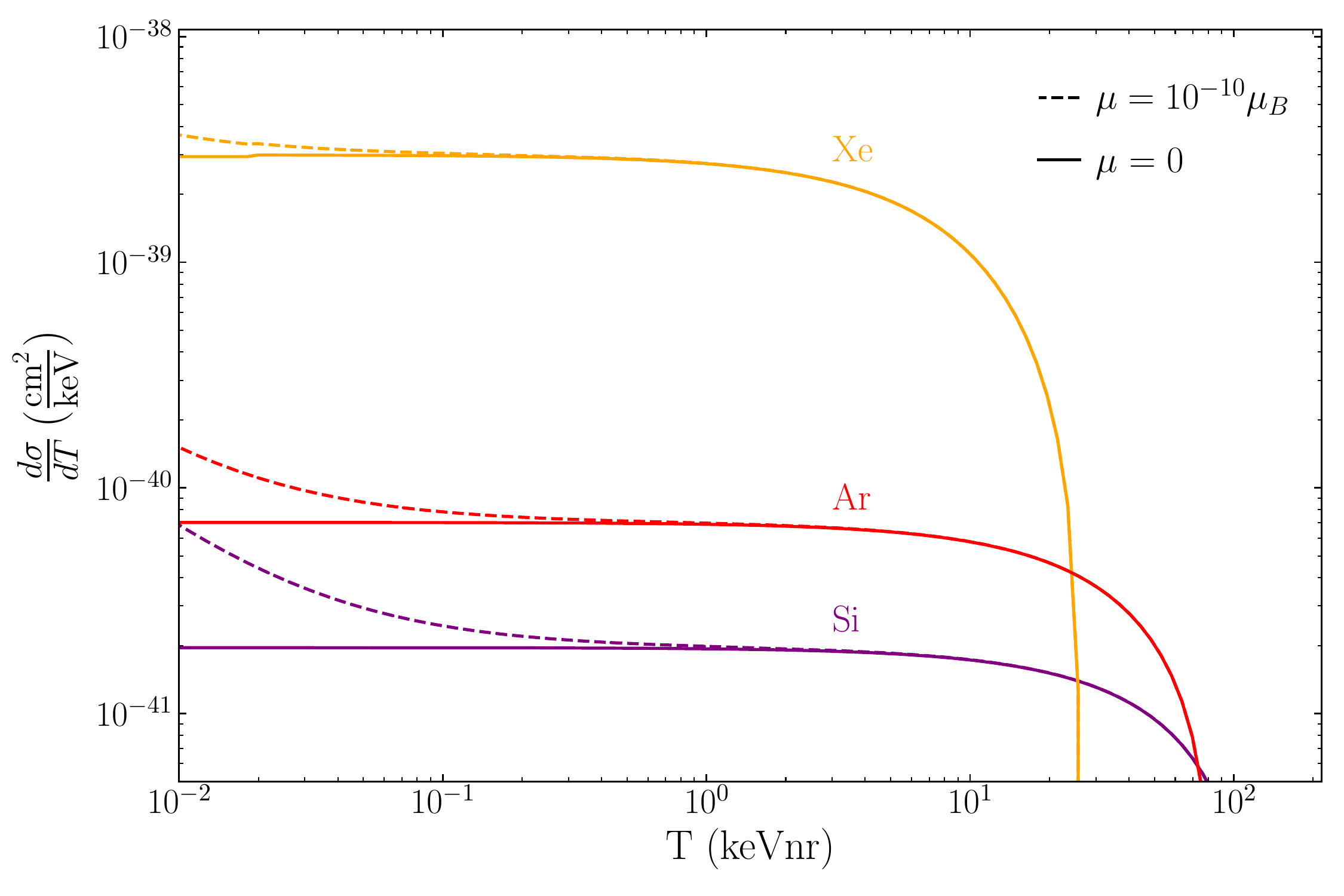}
\caption{\label{fig:xsec-mu} Differential cross section for CE$\nu$NS,
  as a function of the recoiling energy of the nucleus, in the SM and
  in presence of a non-vanishing neutrino magnetic moment $\mu =
  10^{-10}\mu_B$. The different colors correspond to different target
  nuclei, as indicated by the labels. The incident neutrino energy is
  set to $E_\nu = 40$~MeV in all cases, and the cross section in the
  SM ($\mu = 0$) is also shown for comparison. }
\end{figure}
%%%%%%%%%%%%%%%%

Figure~\ref{fig:xsec-mu} shows the differential cross section in
Eq.~\eqref{eq:xsec-mu} as a function of the recoil energy, for an
incident neutrino with energy $E_\nu = 40$~MeV. The different lines
correspond to different target nuclei, as indicated in the legend. As
can be seen from the figure, the effect of a finite neutrino magnetic moment on event rates is noticeable
for recoil energies below $0.5-1~{\rm keV }_{nr}$, depending on the target nucleus
being considered. Therefore, it is expected that  detectors with
the lowest possible recoil energy threshold, like those showcased in this paper, will be most sensitive to this
neutrino property. In this respect, the scintillating bubble chamber and the
charged coupled device stand out among the detectors listed in
Table~\ref{tab:detectors}, with lower thresholds well below those for
the rest of technologies considered.

Our results of the expected sensitivity to neutrino magnetic moments
with the detectors under consideration are
summarized in Table~\ref{tab:macrotable}.
For the Ar bubble chamber, in spite of having the lowest detection threshold,
the lack of energy resolution makes this configuration systematics-limited.  
However, bubble chamber detectors offer the
possibility of adjusting their low energy threshold.  While the event
rates decrease as the detector threshold is increased (see
Fig.~\ref{fig:rates}), splitting the running time of a bubble chamber
into two periods (with different detection thresholds) would lead to
an efficient cancellation of systematic uncertainties, and to a boost in
the sensitivity to this model. 
Therefore, the results shown in Table~\ref{tab:macrotable} for this
detector are obtained for a configuration where the data taking time is
split evenly between two exposures with a different detector threshold:
1.5 years running with $E_{th} = 0.1~$keVnr, and 1.5 years running with
$E_{th}=1$~keVnr. In this case, the systematic uncertainties are taken
to be fully correlated between the two samples. To this end, the
threshold can be trivially alternated between consecutive events, by
varying the operating pressure of the chamber.

Neutrino magnetic moments arise in a variety of models of NP
and, in particular, they do not need to be
flavor-universal. Therefore, being as general as possible, we allow
different magnetic moments for the different neutrino flavors.
However, reactor and solar experiments (among others) bound the
$\nu_e$ magnetic moments at the level of $10^{-11}$ $\mu_{\rm
  B}$~\cite{pdg,Beda:2012zz} and render its effect well beyond the
reach of any of the CE$\nu$NS experiments considered here.  For this
reason we only show on Table~\ref{tab:macrotable} our results on the
sensitivity to $\mu_{\nu_\mu}$ for which current scattering
experiments yield a bound of $6.8\times 10^{-10}$ at 90\%
CL~\cite{Cadeddu:2019eta} which, as seen in the table, can be improved
with some of the considered setups.

\section{Conclusions}
\label{sec:conclusions}
Operating at a proton energy of 2 GeV and 5 MW power, the ESS will
soon become not just the most intense source of spallation neutrons,
but also of pion DAR neutrinos. Its predicted neutrino yield is
unmatched by present or planned spallation sources. As such, it
provides an extensive enhancement in sensitivity to many areas of
phenomenology reachable via CE$\nu$NS experimentation. While we have
not included this possibility in our analysis, it should be noted that
presently contemplated future ESS upgrades include doubling the beam
spill rate to 28 Hz, and/or increasing the proton energy to 2.5-3
GeV. Both options would lead to further improvements in sensitivity.

Recent work \cite{csiqf} has shown that systematic uncertainties
affecting CE$\nu$NS experiments can be reduced to a cumulative
$\sim$10\% level. Realistically, and from a present-day perspective,
further progress in this direction should be considered very hard to
achieve. In this context, the subdominant statistical uncertainty
possible for CE$\nu$NS measurements at the ESS acquires a special
relevance: the experimentation we have described will provide close to
the best possible sensitivity to neutrino properties that can be
expected from CE$\nu$NS studies at spallation sources.

In this work we have explored the sensitivity to a few representative
NP scenarios which illustrate the potential of our proposal
from a quantitative perspective: NSI,
precise determination of the weak mixing angle, neutrino charge
radii and neutrino magnetic moments.  { 
We find that, while flavour discrimination associated to
temporal separation of the signal would not be possible at the ESS, the large
statistics expected still allows for a partial separation between different
neutrino flavors using the energy information}.
On the detector side, we have considered a suite of
innovative detector technologies that originate in the fields of dark
matter detection and double-beta decay searches (listed in
Tab.~\ref{tab:detectors}), and that will allow to maximally profit
from the order-of-magnitude increase in neutrino flux. Specifically,
we find that any of the detectors considered here can lead to an
{ improved precision  of the determination of the}
weak mixing angle in CE$\nu$NS by a factor
$\mathcal{O}(3)$ and to {tighter} bounds on $\nu_\mu$ magnetic moment
by a similar factor. A { substantial enhancement} on the sensitivity
to NSI is also expected as shown in Fig.~\ref{fig:nsi-alldets}.
{ We have also explored the improvement in reach obtained 
from a combination of nuclear targets and quantified this effect in the
context of NSI, see Fig.~\ref{fig:next}}.

Finally, we want to emphasize that our proposal is not premature, as
the time to develop and implement these innovative nuclear recoil
detector technologies is a good match to the start of the ESS users'
program in 2023. By employing a variety of detectors, we expect to
provide an enhanced sensitivity to neutrino properties, and the
ability to confirm or refute any deviations from the { SM}
that might be observed via CE$\nu$NS at the ESS.

\section*{Acknowledgments}

We are indebted to Z. Lazic, R. Linander, M. Lindroos, and V. Santoro
for their input regarding ESS specifics. We also thank O. Borshchev at
Luminnotech for kindly providing us with NOL wavelength shifter
samples, and M. McClish and P. Waer at RMD Inc. for LAAPD-related
consultations. PC warmly thanks B. Dutta for useful discussions.
IE acknowledges support from the FPU program
fellowship FPU15/0369.
PC thanks the CERN Theory division for support
and hospitality during the final stages of this work.
MCGG thanks the
Department of Physics at Columbia University for their hospitality.

This work was supported by the MINECO grants FPA2017-85985-P and
FPA2016-76005-C2-1-P, by PROMETEO/2019/083, by USA-NSF grants
PHY-1620628, PHY-1806722, and PHY-1812702, USA-DARPA award
W911NF1810222, by EU Networks FP10 ITN ELUSIVES
(H2020-MSCA-ITN-2015-674896) and INVISIBLES-PLUS
(H2020-MSCA-RISE-2015-690575), and by AGAUR (Generalitat de Catalunya)
grant 2017-SGR-929. This work was supported in part by the Kavli Institute for Cosmological Physics at the University of Chicago through an endowment from the Kavli Foundation and its founder Fred Kavli.

\bibliographystyle{JHEP}
\bibliography{references}

\providecommand{\noopsort}[1]{}\providecommand{\singleletter}[1]{#1}%
\providecommand{\href}[2]{#2}\begingroup\raggedright\begin{thebibliography}{100}

\bibitem{leo}
A.~Drukier and L.~Stodolsky, {\it {Principles and Applications of a Neutral
  Current Detector for Neutrino Physics and Astronomy}},  {\em Phys. Rev.} {\bf
  D30} (1984) 2295. [,395(1984)].

\bibitem{science}
{\bf COHERENT} Collaboration, D.~Akimov {\em et~al.}, {\it {Observation of
  Coherent Elastic Neutrino-Nucleus Scattering}},  {\em Science} {\bf 357}
  (2017), no.~6356 1123--1126, [\href{http://xxx.lanl.gov/abs/1708.01294}{{\tt
  1708.01294}}].

\bibitem{freedman}
D.~Z. Freedman, {\it {Coherent Neutrino Nucleus Scattering as a Probe of the
  Weak Neutral Current}},  {\em Phys. Rev.} {\bf D9} (1974) 1389--1392.

\bibitem{ournim}
J.~I. Collar, N.~E. Fields, M.~Hai, T.~W. Hossbach, J.~L. Orrell, C.~T.
  Overman, G.~Perumpilly, and B.~Scholz, {\it {Coherent neutrino-nucleus
  scattering detection with a CsI[Na] scintillator at the SNS spallation
  source}},  {\em Nucl. Instrum. Meth.} {\bf A773} (2015) 56--65,
  [\href{http://xxx.lanl.gov/abs/1407.7524}{{\tt 1407.7524}}].

\bibitem{bjorn}
B.~J. Scholz, {\em {First Observation of Coherent Elastic Neutrino-Nucleus
  Scattering}}.
\newblock PhD thesis, Chicago U., Cham, 2017.
\newblock \href{http://xxx.lanl.gov/abs/1904.01155}{{\tt 1904.01155}}.

\bibitem{nsi2}
P.~Coloma, M.~C. Gonzalez-Garcia, M.~Maltoni, and T.~Schwetz, {\it {COHERENT
  Enlightenment of the Neutrino Dark Side}},  {\em Phys. Rev.} {\bf D96}
  (2017), no.~11 115007, [\href{http://xxx.lanl.gov/abs/1708.02899}{{\tt
  1708.02899}}].

\bibitem{nsi1}
J.~B. Dent, B.~Dutta, S.~Liao, J.~L. Newstead, L.~E. Strigari, and J.~W.
  Walker, {\it {Probing light mediators at ultralow threshold energies with
  coherent elastic neutrino-nucleus scattering}},  {\em Phys. Rev.} {\bf D96}
  (2017), no.~9 095007, [\href{http://xxx.lanl.gov/abs/1612.06350}{{\tt
  1612.06350}}].

\bibitem{nsi3}
J.~Liao and D.~Marfatia, {\it {COHERENT constraints on nonstandard neutrino
  interactions}},  {\em Phys. Lett.} {\bf B775} (2017) 54--57,
  [\href{http://xxx.lanl.gov/abs/1708.04255}{{\tt 1708.04255}}].

\bibitem{nsi4}
J.~B. Dent, B.~Dutta, S.~Liao, J.~L. Newstead, L.~E. Strigari, and J.~W.
  Walker, {\it {Accelerator and reactor complementarity in coherent
  neutrino-nucleus scattering}},  {\em Phys. Rev.} {\bf D97} (2018), no.~3
  035009, [\href{http://xxx.lanl.gov/abs/1711.03521}{{\tt 1711.03521}}].

\bibitem{nsi5}
Y.~Farzan, M.~Lindner, W.~Rodejohann, and X.-J. Xu, {\it {Probing neutrino
  coupling to a light scalar with coherent neutrino scattering}},  {\em JHEP}
  {\bf 05} (2018) 066, [\href{http://xxx.lanl.gov/abs/1802.05171}{{\tt
  1802.05171}}].

\bibitem{nsi6}
M.~Abdullah, J.~B. Dent, B.~Dutta, G.~L. Kane, S.~Liao, and L.~E. Strigari,
  {\it {Coherent elastic neutrino nucleus scattering as a probe of a Z′
  through kinetic and mass mixing effects}},  {\em Phys. Rev.} {\bf D98}
  (2018), no.~1 015005, [\href{http://xxx.lanl.gov/abs/1803.01224}{{\tt
  1803.01224}}].

\bibitem{nsi7}
I.~Esteban, M.~C. Gonzalez-Garcia, M.~Maltoni, I.~Martinez-Soler, and
  J.~Salvado, {\it {Updated Constraints on Non-Standard Interactions from
  Global Analysis of Oscillation Data}},  {\em JHEP} {\bf 08} (2018) 180,
  [\href{http://xxx.lanl.gov/abs/1805.04530}{{\tt 1805.04530}}].

\bibitem{nsi8}
D.~Aristizabal~Sierra, V.~De~Romeri, and N.~Rojas, {\it {COHERENT analysis of
  neutrino generalized interactions}},  {\em Phys. Rev.} {\bf D98} (2018)
  075018, [\href{http://xxx.lanl.gov/abs/1806.07424}{{\tt 1806.07424}}].

\bibitem{nsi9}
I.~M. Shoemaker, {\it {COHERENT search strategy for beyond standard model
  neutrino interactions}},  {\em Phys. Rev.} {\bf D95} (2017), no.~11 115028,
  [\href{http://xxx.lanl.gov/abs/1703.05774}{{\tt 1703.05774}}].

\bibitem{Giunti:2019xpr}
C.~Giunti, {\it {General COHERENT Constraints on Neutrino Non-Standard
  Interactions}},  \href{http://xxx.lanl.gov/abs/1909.00466}{{\tt 1909.00466}}.

\bibitem{Denton:2018xmq}
P.~B. Denton, Y.~Farzan, and I.~M. Shoemaker, {\it {Testing large non-standard
  neutrino interactions with arbitrary mediator mass after COHERENT data}},
  {\em JHEP} {\bf 07} (2018) 037,
  [\href{http://xxx.lanl.gov/abs/1804.03660}{{\tt 1804.03660}}].

\bibitem{nst1}
M.~Cadeddu, C.~Giunti, Y.~F. Li, and Y.~Y. Zhang, {\it {Average CsI neutron
  density distribution from COHERENT data}},  {\em Phys. Rev. Lett.} {\bf 120}
  (2018), no.~7 072501, [\href{http://xxx.lanl.gov/abs/1710.02730}{{\tt
  1710.02730}}].

\bibitem{nst2}
E.~Ciuffoli, J.~Evslin, Q.~Fu, and J.~Tang, {\it {Extracting nuclear form
  factors with coherent neutrino scattering}},  {\em Phys. Rev.} {\bf D97}
  (2018), no.~11 113003, [\href{http://xxx.lanl.gov/abs/1801.02166}{{\tt
  1801.02166}}].

\bibitem{nst3}
E.~Ciuffoli, J.~Evslin, Q.~Fu, and J.~Tang, {\it {Extracting nuclear form
  factors with coherent neutrino scattering}},  {\em Phys. Rev.} {\bf D97}
  (2018), no.~11 113003, [\href{http://xxx.lanl.gov/abs/1801.02166}{{\tt
  1801.02166}}].

\bibitem{nst4}
D.~K. Papoulias, T.~S. Kosmas, R.~Sahu, V.~K.~B. Kota, and M.~Hota, {\it
  {Constraining nuclear physics parameters with current and future COHERENT
  data}},  \href{http://xxx.lanl.gov/abs/1903.03722}{{\tt 1903.03722}}.

\bibitem{Cadeddu:2019eta}
M.~Cadeddu, F.~Dordei, C.~Giunti, Y.~F. Li, and Y.~Y. Zhang, {\it {Neutrino,
  Electroweak and Nuclear Physics from COHERENT Elastic Neutrino-Nucleus
  Scattering with Refined Quenching Factor}},
  \href{http://xxx.lanl.gov/abs/1908.06045}{{\tt 1908.06045}}.

\bibitem{em1}
D.~K. Papoulias and T.~S. Kosmas, {\it {COHERENT constraints to conventional
  and exotic neutrino physics}},  {\em Phys. Rev.} {\bf D97} (2018), no.~3
  033003, [\href{http://xxx.lanl.gov/abs/1711.09773}{{\tt 1711.09773}}].

\bibitem{em2}
J.~Billard, J.~Johnston, and B.~J. Kavanagh, {\it {Prospects for exploring New
  Physics in Coherent Elastic Neutrino-Nucleus Scattering}},  {\em JCAP} {\bf
  1811} (2018), no.~11 016, [\href{http://xxx.lanl.gov/abs/1805.01798}{{\tt
  1805.01798}}].

\bibitem{em3}
M.~Cadeddu, C.~Giunti, K.~A. Kouzakov, Y.~F. Li, A.~I. Studenikin, and Y.~Y.
  Zhang, {\it {Neutrino Charge Radii from COHERENT Elastic Neutrino-Nucleus
  Scattering}},  {\em Phys. Rev.} {\bf D98} (2018), no.~11 113010,
  [\href{http://xxx.lanl.gov/abs/1810.05606}{{\tt 1810.05606}}].

\bibitem{em4}
O.~G. Miranda, D.~K. Papoulias, M.~Tórtola, and J.~W.~F. Valle, {\it {Probing
  neutrino transition magnetic moments with coherent elastic neutrino-nucleus
  scattering}},  {\em JHEP} {\bf 07} (2019) 103,
  [\href{http://xxx.lanl.gov/abs/1905.03750}{{\tt 1905.03750}}].

\bibitem{Papoulias:2019txv}
D.~K. Papoulias, {\it {COHERENT constraints after the Chicago-3 quenching
  factor measurement}},  \href{http://xxx.lanl.gov/abs/1907.11644}{{\tt
  1907.11644}}.

\bibitem{wma1}
B.~C. Cañas, E.~A. Garcés, O.~G. Miranda, and A.~Parada, {\it {Future
  perspectives for a weak mixing angle measurement in coherent elastic neutrino
  nucleus scattering experiments}},  {\em Phys. Lett.} {\bf B784} (2018)
  159--162, [\href{http://xxx.lanl.gov/abs/1806.01310}{{\tt 1806.01310}}].

\bibitem{wma2}
M.~Cadeddu and F.~Dordei, {\it {Reinterpreting the weak mixing angle from
  atomic parity violation in view of the Cs neutron rms radius measurement from
  COHERENT}},  {\em Phys. Rev.} {\bf D99} (2019), no.~3 033010,
  [\href{http://xxx.lanl.gov/abs/1808.10202}{{\tt 1808.10202}}].

\bibitem{wma3}
X.-R. Huang and L.-W. Chen, {\it {Neutron Skin in CsI and Low-Energy Effective
  Weak Mixing Angle from COHERENT Data}},  {\em Phys. Rev.} {\bf D100} (2019),
  no.~7 071301, [\href{http://xxx.lanl.gov/abs/1902.07625}{{\tt 1902.07625}}].

\bibitem{ste1}
T.~S. Kosmas, D.~K. Papoulias, M.~Tortola, and J.~W.~F. Valle, {\it {Probing
  light sterile neutrino signatures at reactor and Spallation Neutron Source
  neutrino experiments}},  {\em Phys. Rev.} {\bf D96} (2017), no.~6 063013,
  [\href{http://xxx.lanl.gov/abs/1703.00054}{{\tt 1703.00054}}].

\bibitem{carlos}
C.~Blanco, D.~Hooper, and P.~Machado, {\it {Constraining Sterile Neutrino
  Interpretations of the LSND and MiniBooNE Anomalies with Coherent Neutrino
  Scattering Experiments}},  \href{http://xxx.lanl.gov/abs/1901.08094}{{\tt
  1901.08094}}.

\bibitem{dm1}
S.-F. Ge and I.~M. Shoemaker, {\it {Constraining Photon Portal Dark Matter with
  Texono and Coherent Data}},  {\em JHEP} {\bf 11} (2018) 066,
  [\href{http://xxx.lanl.gov/abs/1710.10889}{{\tt 1710.10889}}].

\bibitem{dm2}
V.~Brdar, W.~Rodejohann, and X.-J. Xu, {\it {Producing a new Fermion in
  Coherent Elastic Neutrino-Nucleus Scattering: from Neutrino Mass to Dark
  Matter}},  {\em JHEP} {\bf 12} (2018) 024,
  [\href{http://xxx.lanl.gov/abs/1810.03626}{{\tt 1810.03626}}].

\bibitem{dm3}
B.~Dutta, D.~Kim, S.~Liao, J.-C. Park, S.~Shin, and L.~E. Strigari, {\it {Dark
  matter signals from timing spectra at neutrino experiments}},
  \href{http://xxx.lanl.gov/abs/1906.10745}{{\tt 1906.10745}}.

\bibitem{diff}
L.~M. Sehgal, {\it {Differences in the Coherent Interactions of $\nu_e$,
  $\nu_\mu$ and $\nu_\tau$}},  {\em Phys. Lett.} {\bf 162B} (1985) 370--372.

\bibitem{essdesign}
R.~Garoby {\em et~al.}, {\it {The European Spallation Source Design}},  {\em
  Phys. Scripta} {\bf 93} (2018), no.~1 014001.

\bibitem{burman}
R.~L. Burman and P.~Plischke, {\it {Neutrino fluxes from a high-intensity
  spallation neutron facility}},  {\em Nucl. Instrum. Meth.} {\bf A398} (1997)
  147--156.

\bibitem{report}
R.L. Burman and P. Plischke, {\it Neutrino Flux Calculations for the proposed
  European Spallation Source}, Forschungszentrum Karlsruhe report FZKA-5834,
  Nov. 1996.

\bibitem{lahet}
\protect{R.\ E.\ Prael, H.\ Lichtenstein, The LAHET Code System, Los Alamos
  National Laboratory report LA-UR-89-30, September 1989.}

\bibitem{louis}
R.~L. Burman and W.~C. Louis, {\it {Neutrino physics at meson factories and
  spallation neutron sources}},  {\em J. Phys.} {\bf G29} (2003) 2499--2512.

\bibitem{mcnpx}
\protect{D.\ Pelowitz {\it et al.}, 2011 MCNPX User's Manual Version 2.7.0
  LA-CP-11-00438}.

\bibitem{g4}
{\bf GEANT4} Collaboration, S.~Agostinelli {\em et~al.}, {\it {GEANT4: A
  Simulation toolkit}},  {\em Nucl. Instrum. Meth.} {\bf A506} (2003) 250--303.

\bibitem{fluka}
T.~T. Böhlen, F.~Cerutti, M.~P.~W. Chin, A.~Fassò, A.~Ferrari, P.~G. Ortega,
  A.~Mairani, P.~R. Sala, G.~Smirnov, and V.~Vlachoudis, {\it {The FLUKA Code:
  Developments and Challenges for High Energy and Medical Applications}},  {\em
  Nucl. Data Sheets} {\bf 120} (2014) 211--214.

\bibitem{essopt}
K.~Batkov, A.~Takibayev, L.~Zanini, and F.~Mezei, {\it Unperturbed moderator
  brightness in pulsed neutron sources},  {\em Nucl. Instr. Meth. Phys. Res. A}
  {\bf 729} (2013) 500 -- 505.

\bibitem{mvsg}
D.~D. DiJulio, K.~Batkov, J.~Stenander, N.~Cherkashyna, and P.~M. Bentley, {\it
  Benchmarking geant4 for spallation neutron source calculations},  {\em J.
  Phys. Conf. Ser.} {\bf 746} (sep, 2016) 012032.

\bibitem{harp}
{\bf HARP} Collaboration, M.~Apollonio {\em et~al.}, {\it {Large-angle
  production of charged pions with incident pion beams on nuclear targets}},
  {\em Phys. Rev.} {\bf C80} (2009) 065207,
  [\href{http://xxx.lanl.gov/abs/0907.1428}{{\tt 0907.1428}}].

\bibitem{harpcdp}
A.~Bolshakova {\em et~al.}, {\it {Cross-sections of large-angle hadron
  production in proton-- and pion--nucleus interactions VIII: aluminium nuclei
  and beam momenta from {$\pm$}3 GeV/c to {$\pm$}15 GeV/c}},  {\em Eur. Phys.
  J.} {\bf C72} (2012) 1882, [\href{http://xxx.lanl.gov/abs/1110.6753}{{\tt
  1110.6753}}].

\bibitem{bench1}
{\bf HARP-CDP} Collaboration, A.~Bolshakova {\em et~al.}, {\it {HARP-CDP
  hadroproduction data: Comparison with FLUKA and GEANT4 simulations}},  {\em
  Eur. Phys. J.} {\bf C70} (2010) 543--553,
  [\href{http://xxx.lanl.gov/abs/1006.3429}{{\tt 1006.3429}}].

\bibitem{bench2}
J.~C. David, {\it {Spallation reactions: A successful interplay between
  modeling and applications}},  {\em Eur. Phys. J.} {\bf A51} (2015), no.~6 68,
  [\href{http://xxx.lanl.gov/abs/1505.03282}{{\tt 1505.03282}}].

\bibitem{bench3}
D.~Mancusi {\em et~al.}, {\it {On the role of secondary pions in spallation
  targets}},  {\em Eur. Phys. J.} {\bf A53} (2017), no.~5 80,
  [\href{http://xxx.lanl.gov/abs/1603.05453}{{\tt 1603.05453}}].

\bibitem{rec1}
{\bf HARP-CDP Group} Collaboration, A.~Bolshakova {\em et~al.}, {\it
  {Revisiting the 'LSND anomaly' I: impact of new data}},  {\em Phys. Rev.}
  {\bf D85} (2012) 092008, [\href{http://xxx.lanl.gov/abs/1110.4265}{{\tt
  1110.4265}}].

\bibitem{jason}
J. Newby's talk at ``The Magnificent CE$\nu$NS'' workshop, November 2-3, 2018,
  University of Chicago. \\
  \url{https://kicp-workshops.uchicago.edu/2018-CEvNS/program.php}.

\bibitem{yurinue}
Y. Efremenko's talk at NuEclipse Workshop, 20-22 August 2017, University of
  Tennessee. \url{http://www.phys.utk.edu/news/2017/nuclipse.html}.

\bibitem{d2o}
{\bf COHERENT} Collaboration, R.~Rapp, {\it {COHERENT Plans for D$_2$O at the
  Spallation Neutron Source}},  in {\em {Meeting of the Division of Particles
  and Fields of the American Physical Society (DPF2019) Boston, Massachusetts,
  July 29-August 2, 2019}}, 2019.
\newblock \href{http://xxx.lanl.gov/abs/1910.00630}{{\tt 1910.00630}}.

\bibitem{scholberg}
K. Scholberg's talk at TAUP 2019, 8-14 September 2019, Toyama.
  \url{http://www-kam2.icrr.u-tokyo.ac.jp/indico/event/3/program}.

\bibitem{rex}
M.~R. Heath.
\newblock PhD thesis, Indiana University, 2019.

\bibitem{rex2}
{\bf COHERENT collaboration} Collaboration, D.~Akimov {\em et~al.}, {\it {First
  Constraint on Coherent Elastic Neutrino-Nucleus Scattering in Argon}},
  \href{http://xxx.lanl.gov/abs/1909.05913}{{\tt 1909.05913}}.

\bibitem{amsler}
C.~Amsler, D.~Grogler, W.~Joffrain, D.~Lindelof, M.~Marchesotti,
  P.~Niederberger, H.~Pruys, C.~Regenfus, P.~Riedler, and A.~Rotondi, {\it
  {Temperature dependence of pure CsI: Scintillation light yield and decay
  time}},  {\em Nucl. Instrum. Meth.} {\bf A480} (2002) 494--500.

\bibitem{mos1}
M.~Moszynski, M.~Balcerzyk, W.~Czarnacki, M.~Kapusta, W.~Klamra, P.~Schotanus,
  A.~Syntfeld, M.~SzawÇowski, and V.~Kozlov, {\it Energy resolution and
  non-proportionality of the light yield of pure csi at liquid nitrogen
  temperatures},  {\em Nucl. Instr. Meth. Phys. Res. A} {\bf 537} (01, 2005)
  357--362.

\bibitem{mos2}
M.~Moszynski, W.~Czarnacki, W.~Klamra, M.~Szawlowski, P.~Schotanus, and
  M.~Kapusta, {\it Application of large area avalanche photodiodes to study
  scintillators at liquid nitrogen temperatures},  {\em Nucl. Instr. Meth.
  Phys. Res. A} {\bf 504} (2003), no.~1 307 -- 312.

\bibitem{nadeau}
P.~Nadeau.
\newblock PhD thesis, Queen's University, 2015.

\bibitem{clark}
M.~Clark, P.~Nadeau, S.~Hills, C.~Dujardin, and P.~Di~Stefano, {\it {Particle
  detection at cryogenic temperatures with undoped CsI}},  {\em Nucl. Instrum.
  Meth.} {\bf A901} (2018) 6--13,
  [\href{http://xxx.lanl.gov/abs/1709.04020}{{\tt 1709.04020}}].

\bibitem{liu}
J.~Liu, M.~Yamashita, and A.~K. Soma, {\it {Light yield of an undoped CsI
  crystal coupled directly to a photomultiplier tube at 77 Kelvin}},  {\em
  JINST} {\bf 11} (2016), no.~10 P10003,
  [\href{http://xxx.lanl.gov/abs/1608.06278}{{\tt 1608.06278}}].

\bibitem{woody}
C.~L. Woody, P.~W. Levy, J.~A. Kierstead, T.~Skwarnicki, Z.~Sobolewski,
  M.~Goldberg, N.~Horwitz, P.~Souder, and D.~F. Anderson, {\it {Readout
  techniques and radiation damage of undoped cesium iodide}},  {\em IEEE Trans.
  Nucl. Sci.} {\bf 37} (1990) 492--499.

\bibitem{zhang}
X.~Zhang {\em et~al.} {\em Radiat. Detect. Technol. Methods} {\bf 2} (2018) 15.

\bibitem{mik}
V.~B. Mikhailik, V.~Kapustyanyk, V.~Tsybulskyi, V.~Rudyk, and H.~Kraus, {\it
  {Luminescence and scintillation properties of CsI -- a potential cryogenic
  scintillator}},  {\em Phys. Status Solidi} {\bf B252} (2015) 804,
  [\href{http://xxx.lanl.gov/abs/1411.6246}{{\tt 1411.6246}}].

\bibitem{csiisspecial}
S.~S. {Gridin}, A.~N. {Belsky}, N.~V. {Shiran}, and A.~V. {Gektin}, {\it
  Channels of energy losses and relaxation in csi:a scintillators (${\rm
  a}={\rm tl}$, in)},  {\em IEEE Trans. Nucl. Sci.} {\bf 61} (Feb, 2014)
  246--251.

\bibitem{maximumLY}
P.~Dorenbos, {\it Light output and energy resolution of ce3+-doped
  scintillators},  {\em Nucl. Instr. Meth. Phys. Res. A} {\bf 486} (2002),
  no.~1 208 -- 213.

\bibitem{angloher}
G.~Angloher {\em et~al.}, {\it {A CsI low temperature detector for dark matter
  search}},  {\em Astropart. Phys.} {\bf 84} (2016) 70--77,
  [\href{http://xxx.lanl.gov/abs/1602.08884}{{\tt 1602.08884}}].

\bibitem{bandgap1}
C.~K. Ong, K.~S. Song, R.~Monnier, and A.~M. Stoneham {\em J. Phys. C: Sol.
  State Phys.} {\bf 12} (nov, 1979) 4641--4646.

\bibitem{csiqf}
J.~I. Collar, A.~R.~L. Kavner, and C.~M. Lewis, {\it {Response of CsI[Na] to
  Nuclear Recoils: Impact on Coherent Elastic Neutrino-Nucleus Scattering
  (CE$\nu$NS)}},  {\em Phys. Rev.} {\bf D100} (2019), no.~3 033003,
  [\href{http://xxx.lanl.gov/abs/1907.04828}{{\tt 1907.04828}}].

\bibitem{rmd}
\protect{Radiation Monitoring Devices, RMD Inc.\, Watertown, MA.}

\bibitem{belle}
Y.~Jin, H.~Aihara, O.~V. Borshchev, D.~A. Epifanov, S.~A. Ponomarenko, and
  N.~M. Surin, {\it {Study of a pure CsI crystal readout by APD for Belle II
  end cap ECL upgrade}},  {\em Nucl. Instrum. Meth.} {\bf A824} (2016)
  691--692.

\bibitem{jin}
Y.~Jin Master's thesis, University of Tokyo, 2015.

\bibitem{saturation}
N.~Osakabe, J.~Endo, A.~Tonomura, T.~Urakami, S.~Ohsuka, H.~Tsuchiya,
  Y.~Tsuchiya, and T.~Kodama, {\it Saturation of multiplication mechanism in
  silicon avalanche photodiodes used for single electron detection},  {\em Rev.
  Sci. Instrum.} {\bf 69} (1998), no.~8 2898--2901.

\bibitem{yang}
L.~Yang, S.~N. Dzhosyuk, J.~M. Gabrielse, P.~R. Huffman, C.~E.~H. Mattoni,
  S.~E. Maxwell, D.~N. McKinsey, and J.~M. Doyle, {\it {Performance of a
  large-area avalanche photodiode at low temperature for scintillation
  detection}},  {\em Nucl. Instrum. Meth.} {\bf A508} (2003) 388--393.

\bibitem{pmthandbook}
A.~Wright, {\em The photomultiplier handbook}.
\newblock Oxford University Press, 2017.

\bibitem{lumin}
\url{https://www.luminnotech.com/products}.

\bibitem{exo}
R.~Neilson {\em et~al.}, {\it {Characterization of large area APDs for the
  EXO-200 detector}},  {\em Nucl. Instrum. Meth.} {\bf A608} (2009) 68--75,
  [\href{http://xxx.lanl.gov/abs/0906.2499}{{\tt 0906.2499}}].

\bibitem{cryopmt}
Specification sheet for R8520-406/R8520-506 PMTs. \\
  \url{https://www.hamamatsu.com}.

\bibitem{migdal}
M.~Ibe, W.~Nakano, Y.~Shoji, and K.~Suzuki, {\it {Migdal Effect in Dark Matter
  Direct Detection Experiments}},  {\em JHEP} {\bf 03} (2018) 194,
  [\href{http://xxx.lanl.gov/abs/1707.07258}{{\tt 1707.07258}}].

\bibitem{nol}
S.~A. Ponomarenko {\em et~al.}, {\it Nanostructured organosilicon luminophores
  and their application in highly efficient plastic scintillators},  {\em
  Nature Sci. Rep.} {\bf 4} (Oct, 2014) 6549 EP --.

\bibitem{nol2}
T.~Y. Starikova, N.~M. Surin, O.~V. Borshchev, S.~A. Pisarev, E.~A. Svidchenko,
  Y.~V. Fedorov, and S.~A. Ponomarenko, {\it A novel highly efficient
  nanostructured organosilicon luminophore with unusually fast
  photoluminescence},  {\em J. Mater. Chem. C} {\bf 4} (2016) 4699--4708.

\bibitem{nol3}
S.~A. Ponomarenko, O.~V. Borshchev, N.~M. Surin, M.~S. Skorotetcky, E.~A.
  Kleymyuk, T.~Y. Starikova, and A.~S. Tereshenko, {\it {Nanostructured
  organosilicon luminophores for efficient and fast elementary particles
  photodetectors}},  in {\em Nanophotonic Materials XIV} (S.~Cabrini,
  G.~Lerondel, A.~M. Schwartzberg, and T.~Mokari, eds.), vol.~10344, pp.~49 --
  58, International Society for Optics and Photonics, SPIE, 2017.

\bibitem{nol4}
S.~A. Ponomarenko, N.~M. Surin, O.~V. Borshchev, M.~S. Skorotetcky, and A.~M.
  Muzafarov, {\it {Nanostructured organosilicon luminophores as a new concept
  of nanomaterials for highly efficient down-conversion of light}},  in {\em
  Nanophotonic Materials XII} (S.~Cabrini, G.~Lerondel, A.~M. Schwartzberg, and
  T.~Mokari, eds.), vol.~9545, pp.~8 -- 16, International Society for Optics
  and Photonics, SPIE, 2015.

\bibitem{nol5}
O.~Borshchev, N.~Surin, M.~Skorotetcky, and S.~Ponomarenko., {\it
  High-efficient optical wavelength shifters: design, properties, application},
   {\em INEOS OPEN} {\bf 2 (4)} (2019) 112.

\bibitem{farrell}
M.~{McClish}, R.~{Farrell}, F.~{Olschner}, M.~R. {Squillante}, G.~{Entine}, and
  K.~S. {Shah}, {\it Characterization of very large silicon avalanche
  photodiodes},  in {\em IEEE Symposium Conference Record Nuclear Science
  2004.}, pp.~1270--1273 Vol. 2, Oct, 2004.

\bibitem{ktev}
{\bf KOTO} Collaboration, Y.~Yanagida and H.~Yoshimoto, {\it {Reusing KTeV CsI
  crystals for J-PARC KOTO experiment}},  {\em PoS} {\bf KAON09} (2009) 021.

\bibitem{damicsnolab}
{\bf DAMIC} Collaboration, A.~Aguilar-Arevalo {\em et~al.}, {\it {Search for
  low-mass WIMPs in a 0.6 kg day exposure of the DAMIC experiment at SNOLAB}},
  {\em Phys. Rev.} {\bf D94} (2016), no.~8 082006,
  [\href{http://xxx.lanl.gov/abs/1607.07410}{{\tt 1607.07410}}].

\bibitem{damicelectron}
{\bf DAMIC} Collaboration, A.~Aguilar-Arevalo {\em et~al.}, {\it {Constraints
  on Light Dark Matter Particles Interacting with Electrons from DAMIC at
  SNOLAB}},  {\em Phys. Rev. Lett.} {\bf 123} (2019) 181802,
  [\href{http://xxx.lanl.gov/abs/1907.12628}{{\tt 1907.12628}}].

\bibitem{sensei}
{\bf SENSEI} Collaboration, O.~Abramoff {\em et~al.}, {\it {SENSEI:
  Direct-Detection Constraints on Sub-GeV Dark Matter from a Shallow
  Underground Run Using a Prototype Skipper-CCD}},  {\em Phys. Rev. Lett.} {\bf
  122} (2019), no.~16 161801, [\href{http://xxx.lanl.gov/abs/1901.10478}{{\tt
  1901.10478}}].

\bibitem{connie}
{\bf CONNIE} Collaboration, A.~Aguilar-Arevalo {\em et~al.}, {\it {Exploring
  low-energy neutrino physics with the Coherent Neutrino Nucleus Interaction
  Experiment (CONNIE)}},  \href{http://xxx.lanl.gov/abs/1906.02200}{{\tt
  1906.02200}}.

\bibitem{damicm}
\url{https://damic.uchicago.edu/index.php}.

\bibitem{priviterataup}
P. Privitera's talk at TAUP 2019, 8-14 September, 2019, Toyama.
  \url{http://www-kam2.icrr.u-tokyo.ac.jp/indico/event/3/}.

\bibitem{esi}
R.~D. {Ryan}, {\it Precision measurements of the ionization energy and its
  temperature variation in high purity silicon radiation detectors},  {\em IEEE
  Trans. Nucl. Sci.} {\bf 20} (Feb, 1973) 473--480.

\bibitem{javier}
{\bf SENSEI} Collaboration, J.~Tiffenberg, M.~Sofo-Haro, A.~Drlica-Wagner,
  R.~Essig, Y.~Guardincerri, S.~Holland, T.~Volansky, and T.-T. Yu, {\it
  {Single-electron and single-photon sensitivity with a silicon Skipper CCD}},
  {\em Phys. Rev. Lett.} {\bf 119} (2017), no.~13 131802,
  [\href{http://xxx.lanl.gov/abs/1706.00028}{{\tt 1706.00028}}].

\bibitem{alvaro}
A.~E. Chavarria {\em et~al.}, {\it {Measurement of the ionization produced by
  sub-keV silicon nuclear recoils in a CCD dark matter detector}},  {\em Phys.
  Rev.} {\bf D94} (2016), no.~8 082007,
  [\href{http://xxx.lanl.gov/abs/1608.00957}{{\tt 1608.00957}}].

\bibitem{photoneutron}
J.~I. Collar, {\it {Applications of an $^{88}Y/Be$ photo-neutron calibration
  source to Dark Matter and Neutrino Experiments}},  {\em Phys. Rev. Lett.}
  {\bf 110} (2013), no.~21 211101,
  [\href{http://xxx.lanl.gov/abs/1303.2686}{{\tt 1303.2686}}].

\bibitem{geccd}
C.~W. {Leitz} {\em et~al.}, {\it {Development of germanium charge-coupled
  devices}},  in {\em High Energy, Optical, and Infrared Detectors for
  Astronomy VIII}, vol.~10709 of {\em Society of Photo-Optical Instrumentation
  Engineers (SPIE) Conference Series}, p.~1070908, July, 2018.

\bibitem{next}
J.~J. Gomez-Cadenas, {\it {Status and prospects of the NEXT experiment for
  neutrinoless double beta decay searches}},  2019.
\newblock \href{http://xxx.lanl.gov/abs/1906.01743}{{\tt 1906.01743}}.

\bibitem{sorensen}
P.~Sorensen, {\it {Electron train backgrounds in liquid xenon dark matter
  search detectors are indeed due to thermalization and trapping}},
  \href{http://xxx.lanl.gov/abs/1702.04805}{{\tt 1702.04805}}.

\bibitem{trapping}
J.~Xu, S.~Pereverzev, B.~Lenardo, J.~Kingston, D.~Naim, A.~Bernstein,
  K.~Kazkaz, and M.~Tripathi, {\it {Electron extraction efficiency study for
  dual-phase xenon dark matter experiments}},  {\em Phys. Rev.} {\bf D99}
  (2019), no.~10 103024, [\href{http://xxx.lanl.gov/abs/1904.02885}{{\tt
  1904.02885}}].

\bibitem{xesurf}
D.~{\relax Yu}. Akimov {\em et~al.}, {\it {First ground-level laboratory test
  of the two-phase xenon emission detector RED-100}},
  \href{http://xxx.lanl.gov/abs/1910.06190}{{\tt 1910.06190}}.

\bibitem{gas_xenon_nuclear}
{\bf NEXT collaboration} Collaboration, J.~Renner {\em et~al.}, {\it
  {Ionization and scintillation of nuclear recoils in gaseous xenon}},  {\em
  Nucl. Instr. Meth. Phys. Res.} {\bf A793} (2015) 62--74,
  [\href{http://xxx.lanl.gov/abs/1409.2853}{{\tt 1409.2853}}].

\bibitem{kr_res}
{\bf NEXT collaboration} Collaboration, G.~Martínez-Lema {\em et~al.}, {\it
  {Calibration of the NEXT-White detector using $^{83m}\mathrm{Kr}$ decays}},
  {\em JINST} {\bf 13} (2018), no.~10 P10014,
  [\href{http://xxx.lanl.gov/abs/1804.01780}{{\tt 1804.01780}}].

\bibitem{next100}
J.~J. Gomez-Cadenas, F.~Monrabal~Capilla, and P.~Ferrario, {\it {High Pressure
  Gas Xenon TPCs for Double Beta Decay Searches}},  {\em Front.in Phys.} {\bf
  7} (2019) 51, [\href{http://xxx.lanl.gov/abs/1903.02435}{{\tt 1903.02435}}].

\bibitem{new}
{\bf NEXT collaboration} Collaboration, F.~Monrabal {\em et~al.}, {\it {The
  Next White (NEW) Detector}},  {\em JINST} {\bf 13} (2018), no.~12 P12010,
  [\href{http://xxx.lanl.gov/abs/1804.02409}{{\tt 1804.02409}}].

\bibitem{ppc}
P.~S. Barbeau, J.~I. Collar, and O.~Tench, {\it {Large-Mass Ultra-Low Noise
  Germanium Detectors: Performance and Applications in Neutrino and
  Astroparticle Physics}},  {\em JCAP} {\bf 0709} (2007) 009,
  [\href{http://xxx.lanl.gov/abs/nucl-ex/0701012}{{\tt nucl-ex/0701012}}].

\bibitem{majorana}
{\bf Majorana} Collaboration, S.~I. Alvis {\em et~al.}, {\it {A Search for
  Neutrinoless Double-Beta Decay in $^{76}$Ge with 26 kg-yr of Exposure from
  the MAJORANA DEMONSTRATOR}},  {\em Phys. Rev.} {\bf C100} (2019), no.~2
  025501, [\href{http://xxx.lanl.gov/abs/1902.02299}{{\tt 1902.02299}}].

\bibitem{gerda}
{\bf GERDA} Collaboration, M.~Agostini {\em et~al.}, {\it {Improved Limit on
  Neutrinoless Double-$\beta$ Decay of $^{76}$Ge from GERDA Phase II}},  {\em
  Phys. Rev. Lett.} {\bf 120} (2018), no.~13 132503,
  [\href{http://xxx.lanl.gov/abs/1803.11100}{{\tt 1803.11100}}].

\bibitem{texono}
H.~T.-K. Wong, {\it {Taiwan EXperiment On NeutrinO — History and Prospects}},
   {\em The Universe} {\bf 3} (2015), no.~4 22--37,
  [\href{http://xxx.lanl.gov/abs/1608.00306}{{\tt 1608.00306}}]. [Int. J. Mod.
  Phys.A33,no.16,1830014(2018)].

\bibitem{conus}
J.~Hakenmüller {\em et~al.}, {\it {Neutron-induced background in the CONUS
  experiment}},  {\em Eur. Phys. J.} {\bf C79} (2019), no.~8 699,
  [\href{http://xxx.lanl.gov/abs/1903.09269}{{\tt 1903.09269}}].

\bibitem{cogent}
{\bf CoGeNT} Collaboration, C.~E. Aalseth {\em et~al.}, {\it {CoGeNT: A Search
  for Low-Mass Dark Matter using p-type Point Contact Germanium Detectors}},
  {\em Phys. Rev.} {\bf D88} (2013) 012002,
  [\href{http://xxx.lanl.gov/abs/1208.5737}{{\tt 1208.5737}}].

\bibitem{cdex}
{\bf CDEX} Collaboration, H.~Jiang {\em et~al.}, {\it {Limits on Light Weakly
  Interacting Massive Particles from the First 102.8 kg ${\times}$ day Data of
  the CDEX-10 Experiment}},  {\em Phys. Rev. Lett.} {\bf 120} (2018), no.~24
  241301, [\href{http://xxx.lanl.gov/abs/1802.09016}{{\tt 1802.09016}}].

\bibitem{gabriela}
G. Ilie's talk at 3rd PIRE-GEMADARC workshop, 5 December, 2018, Knoxville, TN.
  \url{https://wasabi.physics.unc.edu/event/10/}.

\bibitem{radford}
R.~Cooper, D.~Radford, P.~Hausladen, and K.~Lagergren, {\it A novel hpge
  detector for gamma-ray tracking and imaging},  {\em Nucl. Instr. Meth. Phys.
  Res. A} {\bf 665} (2011) 25 -- 32.

\bibitem{plastic}
C.~Awe, P.~S. Barbeau, J.~I. Collar, S.~Hedges, and L.~Li, {\it {Liquid
  scintillator response to proton recoils in the 10–100 keV range}},  {\em
  Phys. Rev.} {\bf C98} (2018), no.~4 045802,
  [\href{http://xxx.lanl.gov/abs/1804.06457}{{\tt 1804.06457}}].

\bibitem{geqf}
B.~J. Scholz, A.~E. Chavarria, J.~I. Collar, P.~Privitera, and A.~E. Robinson,
  {\it {Measurement of the low-energy quenching factor in germanium using an
  $^{88}$Y/Be photoneutron source}},  {\em Phys. Rev.} {\bf D94} (2016), no.~12
  122003, [\href{http://xxx.lanl.gov/abs/1608.03588}{{\tt 1608.03588}}].

\bibitem{chambers}
W.~J. Bolte {\em et~al.}, {\it {Development of bubble chambers with enhanced
  stability and sensitivity to low-energy nuclear recoils}},  {\em Nucl.
  Instrum. Meth.} {\bf A577} (2007) 569--573,
  [\href{http://xxx.lanl.gov/abs/astro-ph/0503398}{{\tt astro-ph/0503398}}].

\bibitem{coupp}
{\bf COUPP} Collaboration, E.~Behnke {\em et~al.}, {\it {Improved
  Spin-Dependent WIMP Limits from a Bubble Chamber}},  {\em Science} {\bf 319}
  (2008) 933--936, [\href{http://xxx.lanl.gov/abs/0804.2886}{{\tt 0804.2886}}].

\bibitem{pico}
{\bf PICO} Collaboration, C.~Amole {\em et~al.}, {\it {Dark Matter Search
  Results from the Complete Exposure of the PICO-60 C$_3$F$_8$ Bubble
  Chamber}},  {\em Phys. Rev.} {\bf D100} (2019), no.~2 022001,
  [\href{http://xxx.lanl.gov/abs/1902.04031}{{\tt 1902.04031}}].

\bibitem{baxter}
{\bf PICO} Collaboration, C.~Amole {\em et~al.}, {\it {Data-Driven Modeling of
  Electron Recoil Nucleation in PICO C$_3$F$_8$ Bubble Chambers}},  {\em Phys.
  Rev.} {\bf D100} (2019) 082006,
  [\href{http://xxx.lanl.gov/abs/1905.12522}{{\tt 1905.12522}}].

\bibitem{eric}
D.~Baxter {\em et~al.}, {\it {First Demonstration of a Scintillating Xenon
  Bubble Chamber for Detecting Dark Matter and Coherent Elastic
  Neutrino-Nucleus Scattering}},  {\em Phys. Rev. Lett.} {\bf 118} (2017),
  no.~23 231301, [\href{http://xxx.lanl.gov/abs/1702.08861}{{\tt 1702.08861}}].

\bibitem{eric2}
E. Dahl's talk at at ``The Magnificent CE$\nu$NS'' workshop, November 2-3,
  2018, University of Chicago. \\
  \url{https://kicp-workshops.uchicago.edu/2018-CEvNS/program.php}.

\bibitem{cirte}
{\bf COUPP} Collaboration, E.~Behnke {\em et~al.}, {\it {Direct Measurement of
  the Bubble Nucleation Energy Threshold in a $CF_3I$ Bubble Chamber}},  {\em
  Phys. Rev.} {\bf D88} (2013) 021101,
  [\href{http://xxx.lanl.gov/abs/1304.6001}{{\tt 1304.6001}}].

\bibitem{nsi10}
P.~Coloma, P.~B. Denton, M.~C. Gonzalez-Garcia, M.~Maltoni, and T.~Schwetz,
  {\it {Curtailing the Dark Side in Non-Standard Neutrino Interactions}},  {\em
  JHEP} {\bf 04} (2017) 116, [\href{http://xxx.lanl.gov/abs/1701.04828}{{\tt
  1701.04828}}].

\bibitem{Khan:2019cvi}
A.~N. Khan and W.~Rodejohann, {\it {New Physics from COHERENT Data with
  Improved Quenching Factors}},  \href{http://xxx.lanl.gov/abs/1907.12444}{{\tt
  1907.12444}}.

\bibitem{pdg}
{\bf Particle Data Group} Collaboration, M.~Tanabashi {\em et~al.}, {\it
  {Review of Particle Physics}},  {\em Phys. Rev.} {\bf D98} (2018), no.~3
  030001.

\bibitem{Erler:2017knj}
J.~Erler and R.~Ferro-Hernandez, {\it {Weak Mixing Angle in the Thomson
  Limit}},  {\em JHEP} {\bf 03} (2018) 196,
  [\href{http://xxx.lanl.gov/abs/1712.09146}{{\tt 1712.09146}}].

\bibitem{Horowitz:2003cz}
C.~J. Horowitz, K.~J. Coakley, and D.~N. McKinsey, {\it {Supernova observation
  via neutrino - nucleus elastic scattering in the CLEAN detector}},  {\em
  Phys. Rev.} {\bf D68} (2003) 023005.

\bibitem{Helm:1956zz}
R.~H. Helm, {\it {Inelastic and Elastic Scattering of 187-Mev Electrons from
  Selected Even-Even Nuclei}},  {\em Phys. Rev.} {\bf 104} (1956) 1466--1475.

\bibitem{javier2}
F.~Izraelevitch {\em et~al.}, {\it {A Measurement of the Ionization Efficiency
  of Nuclear Recoils in Silicon}},  {\em JINST} {\bf 12} (2017), no.~06 P06014,
  [\href{http://xxx.lanl.gov/abs/1702.00873}{{\tt 1702.00873}}].

\bibitem{Weinberg:1979sa}
S.~Weinberg, {\it {Baryon and Lepton Nonconserving Processes}},  {\em Phys.
  Rev. Lett.} {\bf 43} (1979) 1566--1570.

\bibitem{Coloma:2016gei}
P.~Coloma and T.~Schwetz, {\it {Generalized mass ordering degeneracy in
  neutrino oscillation experiments}},  {\em Phys. Rev.} {\bf D94} (2016), no.~5
  055005, [\href{http://xxx.lanl.gov/abs/1604.05772}{{\tt 1604.05772}}].
  [Erratum: Phys. Rev.D95,no.7,079903(2017)].

\bibitem{Gonzalez-Garcia:2013usa}
M.~C. Gonzalez-Garcia and M.~Maltoni, {\it {Determination of matter potential
  from global analysis of neutrino oscillation data}},  {\em JHEP} {\bf 09}
  (2013) 152, [\href{http://xxx.lanl.gov/abs/1307.3092}{{\tt 1307.3092}}].

\bibitem{Bakhti:2014pva}
P.~Bakhti and Y.~Farzan, {\it {Shedding light on LMA-Dark solar neutrino
  solution by medium baseline reactor experiments: JUNO and RENO-50}},  {\em
  JHEP} {\bf 07} (2014) 064, [\href{http://xxx.lanl.gov/abs/1403.0744}{{\tt
  1403.0744}}].

\bibitem{Barranco:2005yy}
J.~Barranco, O.~G. Miranda, and T.~I. Rashba, {\it {Probing new physics with
  coherent neutrino scattering off nuclei}},  {\em JHEP} {\bf 12} (2005) 021,
  [\href{http://xxx.lanl.gov/abs/hep-ph/0508299}{{\tt hep-ph/0508299}}].

\bibitem{Zeller:2001hh}
{\bf NuTeV} Collaboration, G.~P. Zeller {\em et~al.}, {\it {A Precise
  determination of electroweak parameters in neutrino nucleon scattering}},
  {\em Phys. Rev. Lett.} {\bf 88} (2002) 091802,
  [\href{http://xxx.lanl.gov/abs/hep-ex/0110059}{{\tt hep-ex/0110059}}].
  [Erratum: Phys. Rev. Lett.90,239902(2003)].

\bibitem{Dorenbosch:1986tb}
{\bf CHARM} Collaboration, J.~Dorenbosch {\em et~al.}, {\it {Experimental
  Verification of the Universality of $\nu_e$ and $\nu_\mu$ Coupling to the
  Neutral Weak Current}},  {\em Phys. Lett.} {\bf B180} (1986) 303--307.

\bibitem{Farzan:2015doa}
Y.~Farzan, {\it {A model for large non-standard interactions of neutrinos
  leading to the LMA-Dark solution}},  {\em Phys. Lett.} {\bf B748} (2015)
  311--315.

\bibitem{Farzan:2015hkd}
Y.~Farzan and I.~M. Shoemaker, {\it {Lepton Flavor Violating Non-Standard
  Interactions via Light Mediators}},  {\em JHEP} {\bf 07} (2016) 033.

\bibitem{Akimov:2018vzs}
{\bf COHERENT} Collaboration, D.~Akimov {\em et~al.}, {\it {COHERENT
  Collaboration data release from the first observation of coherent elastic
  neutrino-nucleus scattering}},
  \href{http://xxx.lanl.gov/abs/1804.09459}{{\tt 1804.09459}}.

\bibitem{Wood:1997zq}
C.~S. Wood, S.~C. Bennett, D.~Cho, B.~P. Masterson, J.~L. Roberts, C.~E.
  Tanner, and C.~E. Wieman, {\it {Measurement of parity nonconservation and an
  anapole moment in cesium}},  {\em Science} {\bf 275} (1997) 1759--1763.

\bibitem{Dzuba:2012kx}
V.~A. Dzuba, J.~C. Berengut, V.~V. Flambaum, and B.~Roberts, {\it {Revisiting
  parity non-conservation in cesium}},  {\em Phys. Rev. Lett.} {\bf 109} (2012)
  203003, [\href{http://xxx.lanl.gov/abs/1207.5864}{{\tt 1207.5864}}].

\bibitem{Anthony:2005pm}
{\bf SLAC E158} Collaboration, P.~L. Anthony {\em et~al.}, {\it {Precision
  measurement of the weak mixing angle in Moller scattering}},  {\em Phys. Rev.
  Lett.} {\bf 95} (2005) 081601,
  [\href{http://xxx.lanl.gov/abs/hep-ex/0504049}{{\tt hep-ex/0504049}}].

\bibitem{Wang:2014bba}
{\bf PVDIS} Collaboration, D.~Wang {\em et~al.}, {\it {Measurement of parity
  violation in electron–quark scattering}},  {\em Nature} {\bf 506} (2014),
  no.~7486 67--70.

\bibitem{Giunti:2014ixa}
C.~Giunti and A.~Studenikin, {\it {Neutrino electromagnetic interactions: a
  window to new physics}},  {\em Rev. Mod. Phys.} {\bf 87} (2015) 531,
  [\href{http://xxx.lanl.gov/abs/1403.6344}{{\tt 1403.6344}}].

\bibitem{Degrassi:1989ip}
G.~Degrassi, A.~Sirlin, and W.~J. Marciano, {\it {Effective Electromagnetic
  Form-factor of the Neutrino}},  {\em Phys. Rev.} {\bf D39} (1989) 287--294.

\bibitem{Vogel:1989iv}
P.~Vogel and J.~Engel, {\it {Neutrino Electromagnetic Form-Factors}},  {\em
  Phys. Rev.} {\bf D39} (1989) 3378.

\bibitem{Kouzakov:2017hbc}
K.~A. Kouzakov and A.~I. Studenikin, {\it {Electromagnetic properties of
  massive neutrinos in low-energy elastic neutrino-electron scattering}},  {\em
  Phys. Rev.} {\bf D95} (2017), no.~5 055013,
  [\href{http://xxx.lanl.gov/abs/1703.00401}{{\tt 1703.00401}}]. [Erratum:
  Phys. Rev.D96,no.9,099904(2017)].

\bibitem{Bernabeu:2000hf}
J.~Bernabeu, L.~G. Cabral-Rosetti, J.~Papavassiliou, and J.~Vidal, {\it {On the
  charge radius of the neutrino}},  {\em Phys. Rev.} {\bf D62} (2000) 113012,
  [\href{http://xxx.lanl.gov/abs/hep-ph/0008114}{{\tt hep-ph/0008114}}].

\bibitem{Bernabeu:2002nw}
J.~Bernabeu, J.~Papavassiliou, and J.~Vidal, {\it {On the observability of the
  neutrino charge radius}},  {\em Phys. Rev. Lett.} {\bf 89} (2002) 101802,
  [\href{http://xxx.lanl.gov/abs/hep-ph/0206015}{{\tt hep-ph/0206015}}].
  [Erratum: Phys. Rev. Lett.89,229902(2002)].

\bibitem{Bernabeu:2002pd}
J.~Bernabeu, J.~Papavassiliou, and J.~Vidal, {\it {The Neutrino charge radius
  is a physical observable}},  {\em Nucl. Phys.} {\bf B680} (2004) 450--478,
  [\href{http://xxx.lanl.gov/abs/hep-ph/0210055}{{\tt hep-ph/0210055}}].

\bibitem{Beda:2012zz}
A.~G. Beda, V.~B. Brudanin, V.~G. Egorov, D.~V. Medvedev, V.~S. Pogosov, M.~V.
  Shirchenko, and A.~S. Starostin, {\it {The results of search for the neutrino
  magnetic moment in GEMMA experiment}},  {\em Adv. High Energy Phys.} {\bf
  2012} (2012) 350150.

\end{thebibliography}\endgroup

\end{document}